\documentclass[aps,pre,twocolumn,reprint,superscriptaddress,groupedaddress,longbibliography]{revtex4-2} 
                                                                                           
\usepackage{amsmath,amssymb,amsfonts,mathrsfs}
\usepackage{amsthm}
\usepackage{graphicx}
\usepackage{xcolor}
\usepackage{bm}
\usepackage{hyperref}
\usepackage{booktabs}
\usepackage{physics}
\usepackage{CJK} %% 20260812
\hypersetup{colorlinks=true,linkcolor=blue,citecolor=blue,urlcolor=blue, 
bookmarks=true,bookmarksnumbered=true,unicode=true, 
pdftitle={Diffusion under competing bulk and surface stopping mechanisms}, 
pdfauthor={Yilin Ye}}

\newcommand{\Var}{\operatorname{Var}}

\newcommand{\expect}[1]{\langle #1 \rangle}

\newcommand{\mrm}[1]{\mathrm{#1}}
\newcommand{\md}{\mrm{d}}
\newcommand{\X}{\mathbf{X}}
\newcommand{\x}{\bm{x}}
\newcommand{\E}{\mathbb{E}}
\newcommand{\p}{\mathbb{P}}

\begin{document}
%\begin{CJK*}{UTF8}{gbsn}
\title{Diffusion under competing bulk and surface stopping mechanisms}
%\author{Yilin Ye (叶依林)}
\author{Yilin Ye}
\email{yyl.libri@gmail.com}
\affiliation{Laboratoire de Physique de la Mati\`ere Condens\'ee, CNRS -- École Polytechnique, Institut Polytechnique de Paris, Palaiseau, 91120, France}                        
\date{\today}

\begin{abstract}
We investigate reflected diffusion in a bounded domain subject to two independent, competing stopping mechanisms: an exponentially distributed bulk lifetime of rate $p$ and a surface reaction triggered when the boundary local time exceeds an independent exponential threshold of rate $q$. 
Denoting by $T$ the stopping time and by $L$ the acquired boundary local time at stopping, 
we derive their marginal distributions, joint Laplace transform, and complete hierarchy of mixed moments. 
These statistics are determined by the splitting probability $\phi$ that surface reaction occurs before bulk decay. 
In particular, we establish the identity $p\expect{T}+q\expect{L}=1$ and show that the cumulative risk $pT+qL$ is exponentially distributed with unit rate. 
We further obtain equivalent representations of $\phi$ in terms of the Robin-Laplacian and the generalized Steklov spectra. Explicit results for a three-dimensional ball reveal how competing rates $p,q$ control $\phi$ and the $(T,L)$ statistics.
Monte Carlo simulations test the universal cumulative-risk law. 
\end{abstract}                                   
\maketitle                                                                            
                                                                                       
\section{Introduction}\label{sec:intro}                                               

Diffusing molecules in confined media often face competing fates. They may be deactivated in the bulk after a finite lifetime, or react on a partially reactive interface after accumulating sufficient contact with it. 
Their competition defines a natural two-clock problem in which spontaneous bulk decay and surface reaction act on the same diffusing trajectory.
Such competition may arise when a bioactive molecule is removed in solution while searching for a localized or membrane-bound target, as in receptor capture or DNA target search \cite{Berg77, lauffenburger1996receptors, halford2004site}. The competing removal channel is idealized as spatially uniform first-order decay. Analogous bulk/surface competition can also arise in catalytic or confined reaction systems \cite{Grebenkov23Molecules}.

At the macroscopic level, the bulk decay of diffusing particles is conventionally described by 
\begin{equation}
\partial_t c(\x,t) = D \Delta c(\x,t)-p \, c(\x,t) \quad (\x \in \Omega), 
\label{eq:macrocp}
\end{equation}
where $c(\x,t)$ is the particle concentration, $D$ is the constant diffusion coefficient, and $p$ is the rate constant of a first-order reaction in the bulk. 
For the spatially uniform concentration $c(t)$, Eq.~\eqref{eq:macrocp} reduces to $\md c(t)/\md t=-p \, c(t)$, and thus $c(t)=c(0)e^{-pt}$. 
Surface reactions are usually described by a proper boundary condition \cite{Grebenkov23Molecules, Grebenkov24Molecules}: a Dirichlet condition for a perfectly absorbing surface, a Robin condition for a partially reactive surface, and a Neumann condition for an inert surface. 
The first-passage time (FPT) of a diffusing particle to a target boundary has been extensively studied, with applications spanning chemical kinetics, biology, and materials science \cite{Redner2001, Meyer11, Mattos12, Bray13, Holcman2013, Benichou14, Metzler2014, Grebenkov16, Grebenkov2019}. 
In this setting, the reactant concentration vanishes on the perfectly absorbing boundary (Dirichlet boundary condition), i.e., $c(\x,t)=0$ for $\x \in \partial\Omega$. 
Furthermore, the Robin boundary condition balances diffusive and reactive fluxes \cite{Collins49, Sano79, grebenkov2006partially, Grebenkov18d, Guerin21, Bressloff22PRE}, i.e.,  
\begin{equation}
-D \, \partial_n c(\x,t) = \kappa \, c(\x,t) \quad (\x \in \partial\Omega) , 
\label{eq:RobinBC}
\end{equation}
where $\partial_n$ is the outward normal derivative and $\kappa \in (0,\infty)$ is the surface reactivity. 
The limits $\kappa \to \infty$ and $\kappa \to 0$ recover the Dirichlet and Neumann boundary conditions, respectively.

Microscopically, a constant first-order rate $p$ means that a surviving particle reacts during $(t,t+\md t)$ with conditional probability $p \, \md t$. Its lifetime $\delta$ is therefore exponentially distributed, i.e., $\p(\delta>t)=e^{-pt}$, or equivalently $\delta\sim\mrm{Exp}(p)$. 
The encounter-based approach \cite{Grebenkov20PRE, Grebenkov20PRL, grebenkov2024encounter} provides a probabilistic representation of the Robin boundary condition. The boundary local time (BLT) $\ell_t$ quantifies cumulative residence of the particle in a thin boundary layer, and a reaction event occurs when $\ell_t$ exceeds an independent random threshold $\hat{\ell}$. A constant reactivity $q= \kappa / D$ with units of inverse length corresponds to an exponentially distributed threshold $\hat\ell \sim \mrm{Exp}(q)$ \cite{Grebenkov20PRL}. This probabilistic formulation decouples the diffusive dynamics (governing $\ell_t$) from the chemical kinetics (governing $\hat{\ell}$), enabling the study of much more general surface reaction mechanisms \cite{Grebenkov21JPA, Benkhadaj22}. 
Meanwhile, the boundary local time has also been applied to transport across semipermeable membranes \cite{Bressloff22c, Bressloff23PRE}. 
The boundary local time and its spectral characterization via the Dirichlet-to-Neumann operator have been thoroughly investigated \cite{Grebenkov2019}.

\begin{figure}[t!]                                                                  
\includegraphics[width=\linewidth]{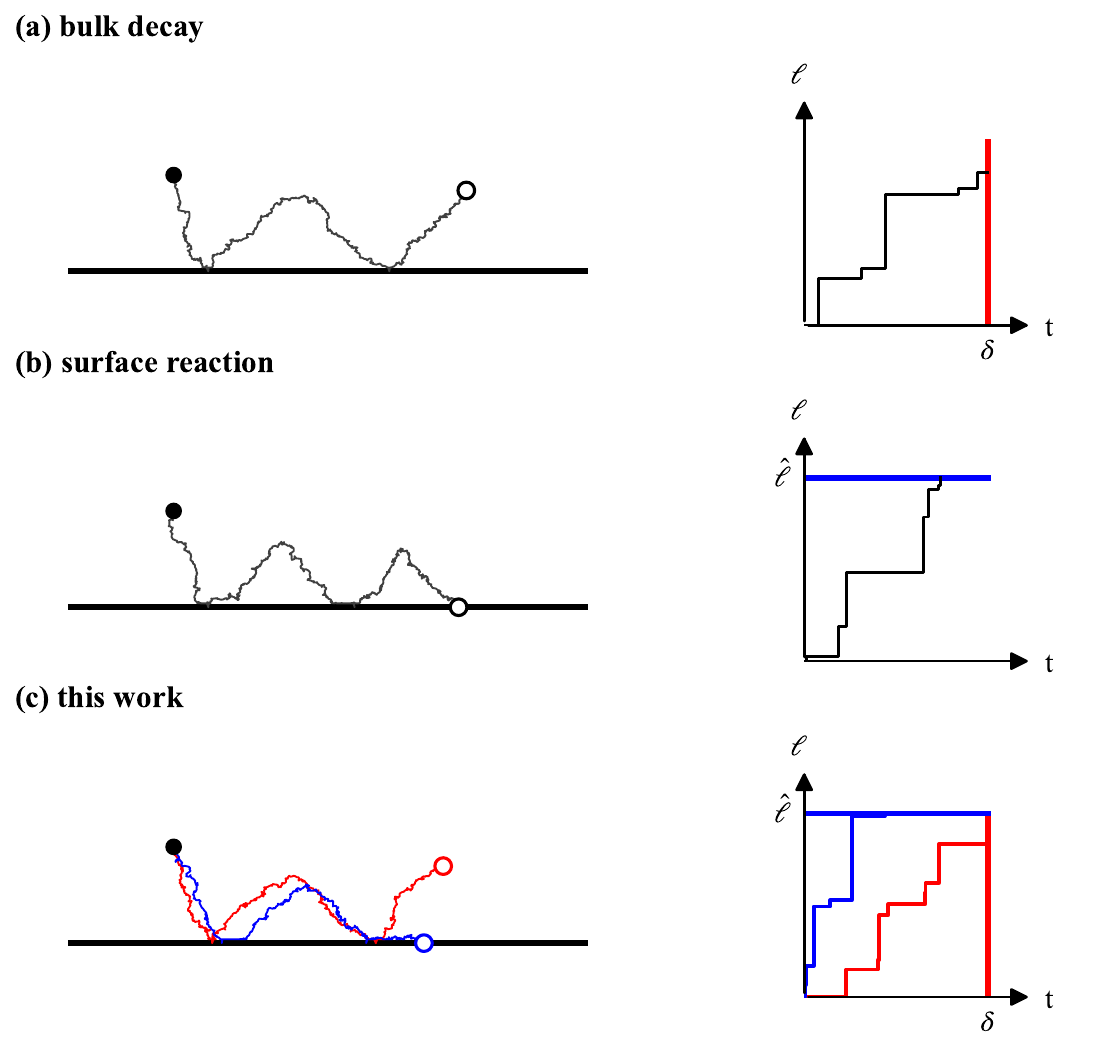}                                 
\caption{
Coarse schematic of the competing stopping mechanisms.
Two independent exponential clocks terminate the trajectory: the bulk lifetime $\delta \sim \mrm{Exp}(p)$ and surface reaction threshold $\hat\ell \sim \mrm{Exp}(q)$. 
\textbf{(a)} Pure bulk decay $(p>0, q=0)$: the trajectory ends at $t = \delta$ (vertical line in red). 
\textbf{(b)} Pure surface reaction $(p=0, q>0)$, the trajectory ends at $\ell = \hat{\ell}$ (horizontal line in blue). 
\textbf{(c)} Competing mechanisms $(p>0,q>0)$: the trajectory may end at either threshold. 
\label{fig:sch}} 
\end{figure}

At the level of the concentration or survival probability, bulk decay and partial surface reactivity can be combined within a reaction–diffusion problem (Eqs.~\eqref{eq:macrocp} and \eqref{eq:RobinBC}). Probabilistic representations of Robin boundary-value problems and spectral descriptions of imperfect reactions are also well established \cite{Papanicolaou90, Zhou16, grebenkov2019spectral, bressloff2022spectral}. 
For the case $p>0$ and $q \to \infty$, related probabilistic studies have treated mortality competing with first passage to perfectly absorbing targets \cite{Yuste13, campos2015optimal, Meerson15, Grebenkov17d, Balakrishnan19, radice2023effects}. 
In the absence of bulk decay, i.e., $p=0$ and $q>0$, distinct boundary local times have been studied for competing surface reactions \cite{Grebenkov20JSM}, while the joint statistics of the first-reaction time and the accumulated boundary local time have been investigated for a single surface-reaction mechanism \cite{Ye26JCP}. 
Through the encounter-based approach, double Laplace transforms with respect to time and local time provide a connection to Robin boundary-value problems \cite{bressloff2022spectral}. The joint statistics of escape time and boundary local time has been characterized in \cite{Grebenkov23PRE}, where the exponential bulk mortality can be incorporated, including the distribution of local time at the bulk death event. However, to our knowledge, the exact theory of the pair $(T,L)$ governed by an independent bulk lifetime and a boundary local time threshold has not been developed.

Building on these previous results, we investigate the joint statistics of the stopped pair $(T,L)$ when an independent exponential bulk lifetime $\delta$ competes with an exponential boundary local time threshold $\hat{\ell}$ along the same reflected trajectory. 
In this paper, we present a comprehensive theoretical framework for diffusion under joint stopping conditions. 
The mathematical structure is remarkably compact: all statistics are reduced to the single fundamental quantity $\phi$, which is the Laplace transform of the first-reaction time distribution under pure Robin conditions evaluated at the bulk decay rate. Once the function $\phi$ is known, the generating function and moments of arbitrary order follow. 
At first order, the probability equivalence shows $\phi = q\expect{L}$, and we obtain the exact sum rule $p\expect{T} + q\expect{L} = 1$ for arbitrary geometries. 
In turn, a more universal law holds: $pT + qL \sim \mrm{Exp}(1)$. 
All mixed moments can be obtained through the generating function. 
Finally, the spectral expansion for $\phi$ leads to a duality between the Robin-Laplacian and the generalized Steklov representations \cite{Levitin2023}.

The paper is organized as follows. 
Section~\ref{sec:theory} presents the theoretical framework: the sum rule of $T$ and $L$, the joint generating function, and the spectral duality. 
Section~\ref{sec:example} provides explicit solutions for the three-dimensional ball and the Monte Carlo validation. 
Section~\ref{sec:conclusion} summarizes the main findings and discusses future perspectives.

\section{Main results}\label{sec:theory}

Consider reflected Brownian motion $\X_t$ with diffusion coefficient $D$ in a bounded domain $\Omega\subset\mathbb R^d$ with smooth and reactive boundary $\partial\Omega = \Gamma$. Note that $\ell_t$ has units of length. 
The particle is subject to two independent stopping mechanisms: (i) a random lifetime $\delta \sim \mrm{Exp}(p)$ with $\p(\delta > t) = e^{-pt}$ ($p > 0$), and (ii) a random boundary local time (BLT) threshold $\hat{\ell} \sim \mrm{Exp}(q)$ with $\p(\hat{\ell} > \ell) = e^{-q\ell}$ ($q > 0$). 
These two random variables $\delta$ and $\hat\ell$ are mutually independent and independent of the reflected Brownian motion.
Their probability density functions (PDFs) are $pe^{-pt}$ and $qe^{-q\ell}$, respectively. 
We denote the first-crossing time (FCT) by $\mathscr{T}_\ell = \inf\{t>0: \ell_t>\ell\}$ for a fixed threshold $\ell$, 
and the first-reaction time (FRT) by $\tau = \inf\{t > 0 : \ell_t > \hat{\ell}\}$. 
Since the trajectory ends when either condition is reached, 
the stopping time is $T = \min(\delta, \tau)$, and the acquired boundary local time is $L = \ell_T = \min(\ell_\delta, \hat\ell)$.

\subsection{Sum rule}
\label{sec:sum}

Survival up to time $t$ requires both $\delta > t$ and $\tau > t$. 
Since $\delta$ is independent of the reflected Brownian motion (and hence of $\tau$), starting from $\x_0$, the survival probability of the joint stopping rule $S_T(t|\x_0)$ factorizes as                             
\begin{align}                                                                      
\label{eq:S_factorization}                                                            
S_T(t|\x_0) %&= \mathbb{P}_{\x_0} (T>t) = \p_{\x_0} (\min(\delta, \tau) > t) \nonumber \\ 
&= \p (\delta > t) \p_{\x_0} (\tau > t) = e^{-pt} S_\tau(t|\x_0) ,                             
\end{align}                                                                        
where 
\begin{equation} 
S_\tau(t|\x_0) = \p_{\x_0}(\tau > t) = \int_t^\infty \md t^\prime \, f_\tau(t^\prime|\x_0) ,
\end{equation}
is the Robin survival probability and $f_\tau(t|\x_0)$ is the probability density function of the first-reaction time $\tau$. 
The probability density of the stopping time follows 
\begin{align}                                                                      
\label{eq:f_T_decomp}                                                                 
f_T(t|\x_0) &= - \dv{}{t} S_T(t|\x_0) = - \dv{}{t} \left[ e^{-pt} S_\tau(t|\x_0) \right] \nonumber \\
&= e^{-pt} f_\tau(t|\x_0) + p\,e^{-pt} S_\tau(t|\x_0),                        
\end{align}                                                                        
where the first term $e^{-pt} f_\tau(t|\x_0)$ is the surface-reaction contribution, denoted as $f_{T,\mrm{s}}(t|\x_0)$, and the second term $p e^{-pt} S_\tau(t|\x_0)$ is the bulk-decay contribution, denoted by $f_{T,\mrm{b}}(t|\x_0)$. 
Each of $f_{T,\mrm{b}}(t|\x_0)$ and $f_{T,\mrm{s}}(t|\x_0)$ is given by the probability density of the corresponding stopping mechanism multiplied by the survival probability of the other mechanism.
These two terms integrate to the splitting probabilities $\p_{\x_0}(\tau < \delta)$ and $\p_{\x_0}(\delta < \tau)$: 
during an infinitesimal interval $(t,t+\md t)$, the particle either reacts at the surface without having decayed earlier, with probability $e^{-pt} f_\tau(t|\x_0) \md t$, or decays in the bulk having survived both mechanisms up to time $t$, with probability $p\,e^{-pt} S_\tau(t|\x_0) \md t$.

As the random lifetime $\delta$ truncates the first-reaction time $\tau$, the mean stopping time under the joint condition is expressed in terms of $f_\tau(t|\x_0)$, 
\begin{equation}
\label{eq:Texpect}
\expect{T} (\x_0) %&= \int_0^\infty \md t \, t \, f_T(t|\x_0) = \int_0^\infty \md t \, e^{-pt} S_\tau(t|\x_0) \nonumber \\
= \frac{1}{p} \left[ 1 - \int_0^\infty \md t \, e^{-pt} f_\tau(t|\x_0) \right] .
\end{equation}
See Appendix~\ref{app:int} for calculation details. 
We denote the Laplace transform of the FRT density evaluated at the bulk decay rate $p$ by $\phi(p,q|\x_0)$: 
\begin{align}                                                                      
\label{eq:H_def}
\phi(p,q|\x_0) &= \int_0^\infty \md t \, e^{-pt} f_\tau(t|\x_0) = \E_{\x_0} \left[ e^{-p \tau} \right] \nonumber \\
&= \int_0^\infty \md t \, f_{T,\mrm{s}}(t|\x_0) = \p_{\x_0}(\tau < \delta) .
\end{align}
Therefore, combining Eqs. \eqref{eq:Texpect} and \eqref{eq:H_def}, we obtain
\begin{equation}
\label{eq:pT1H}
p \expect{T}(\x_0) = \p_{\x_0}(\delta<\tau) = 1-\phi(p,q|\x_0) .
\end{equation}

The distribution of the boundary local time $\ell_t$ is known in the Laplace domain \cite{Grebenkov2019, Ye25JCP}. We denote its probability measure by $\Psi(\ell,p|\x_0)$, which contains an atom and a continuous part:
\begin{equation}
\label{eq:BLTPsi}
\Psi(\ell,p|\x_0) = \pi_0(p|\x_0) \delta_0(\ell) + (1-\pi_0) \psi(\ell,p|\x_0) , 
\end{equation}
where $\delta_0(z)$ is the Dirac delta distribution, $\pi_0(p|\x_0)$ is the probability that particles starting from $\x_0$ acquire zero boundary local time and decay with the rate $p$ before their first boundary encounter, and $\psi(\ell,p|\x_0)$ is the PDF of $\ell>0$. 
Indeed, explicit forms of $\Psi(\ell,p|\x_0)$ or $\psi(\ell,p|\x_0)$ are unnecessary since they disappear in the following calculations.  
As there is a singular contribution at $\ell = 0$,  
we denote the survival probability up to $\ell_\delta$ as $Q_\delta(\ell|\x_0)$ for $\ell>0$:  
\begin{equation}
Q_\delta(\ell|\x_0) = \p_{\x_0}(\ell_\delta > \ell ) = \int_\ell^\infty \md \ell^\prime \, \Psi(\ell^\prime,p|\x_0) . 
\end{equation}
Note that $Q_\delta(0^+|\x_0)=1-\pi_0(p|\x_0) \leqslant 1$, where the equality holds for starting points on the reactive boundary. 
Thus, as in Eq.~\eqref{eq:S_factorization}, for the boundary local time $L$ under the joint condition, the two competing clocks yield its survival function $Q_L(\ell|\x_0)$ for $\ell>0$: 
\begin{align}
Q_L(\ell|\x_0) %&= \p_{\x_0}(L>\ell) = \p_{\x_0} (\min(\ell_\delta, \hat\ell) > \ell) \\
&= \p_{\x_0} (\ell_\delta > \ell) \p (\hat\ell > \ell) = e^{-q\ell} Q_\delta(\ell|\x_0) . %\nonumber 
\end{align}

Let $g_L(\ell|\x_0)$ denote the generalized probability density function of $L = \ell_T$ (including a Dirac delta distribution at $\ell=0$): 
\begin{align}
g_L(\ell|\x_0) &= - \dv{}{\ell} Q_L(\ell|\x_0) = - \dv{}{\ell} \left[ e^{-q\ell} Q_\delta(\ell|\x_0) \right] \nonumber \\
&= q e^{-q \ell} Q_\delta(\ell|\x_0) + e^{-q \ell} \Psi(\ell, p |\x_0) .
\end{align}
Again, the first term $g_{L,\mrm{s}}(\ell|\x_0) = q e^{-q \ell} Q_\delta(\ell|\x_0)$ refers to the surface-reaction contribution, while the second term $g_{L,\mrm{b}}(\ell|\x_0) = e^{-q \ell} \Psi(\ell, p |\x_0)$ refers to the bulk-decay contribution.
The mean acquired boundary local time at stopping reads
\begin{equation}
\label{eq:Lexpect}
\expect{L}(\x_0) %&= \int_0^\infty \md \ell \, \ell \, g_L(\ell|\x_0) = \int_0^\infty \md \ell \, e^{-q\ell} Q_\delta (\ell|\x_0) \nonumber \\
= \frac{1}{q} \left[ 1 - \int_0^\infty \md \ell \, e^{-q \ell} \Psi(\ell,p|\x_0) \right] .
\end{equation}
Similarly, see Appendix~\ref{app:int} for calculation details. 
Note that
\begin{align}
\int_0^\infty \md \ell \, g_{L,\mrm{b}}(\ell|\x_0) &= \int_0^\infty \md \ell \, e^{-q \ell} \Psi(\ell,p|\x_0) = \E_{\x_0} \left[ e^{-q\ell_\delta} \right] \nonumber \\
&= \p_{\x_0}(\delta < \tau) = 1 - \phi(p,q|\x_0), 
\end{align}
and we obtain
\begin{align}
\label{eq:qLH}
q \expect{L}(\x_0) &= 1 - \int_0^\infty \md \ell \, e^{-q \ell} \Psi(\ell,p|\x_0) \nonumber \\
&= \int_0^\infty \md \ell \, q \, e^{-q\ell} Q_\delta(\ell|\x_0) = \p_{\x_0}(\hat\ell < \ell_\delta) \nonumber \\
&= \p_{\x_0}(\tau < \delta) = \phi(p,q|\x_0) . 
\end{align}

Since the boundary local time $\ell_t$ is continuous and nondecreasing, $\tau<\delta$ is equivalent to $\hat\ell<\ell_\delta$. For the complementary event, $\delta<\tau$ implies $\ell_\delta \leq \hat\ell$, as equality may occur on a plateau of the boundary local time. 
However, the equality $\hat\ell = \ell_\delta$ (or $\tau = \delta$) has zero probability because $\hat\ell$ has a continuous distribution and is independent of $\ell_\delta$. 
Therefore, according to Eqs. \eqref{eq:pT1H} and \eqref{eq:qLH}, this yields a central identity in this paper: 
\begin{equation} 
\label{eq:sum_rule}
p \expect{T}(\x_0) + q \expect{L}(\x_0) = 1 ,
\end{equation}
which holds for arbitrary $\x_0 \in \Omega$, $p > 0$, and $q > 0$. The interpretation is transparent: 
$q\expect{L}(\x_0) = \p_{\x_0}(\tau < \delta) = \phi(p,q|\x_0)$ is the probability of surface-triggered stopping, 
while $p\expect{T}(\x_0) = \p_{\x_0}(\delta < \tau) = 1 - \phi(p,q|\x_0)$ is the effective probability of lifetime-triggered stopping. 
Their sum is unity because one of the two mechanisms must eventually stop the particle.

Although the formula \eqref{eq:Texpect} appears singular as $p \to 0$, this apparent singularity is removable. The mean stopping time is still finite: 
\begin{equation}                                                                      
\label{eq:Tlimit}                                                                     
\lim_{p \to 0} \expect{T}(\x_0) = \lim_{p \to 0} \frac{1-\phi(p,q|\x_0)}{p} = -\frac{\partial \phi(p,q|\x_0)}{\partial p}\bigg|_{p=0},               
\end{equation}
which recovers the classical mean first-reaction time under the Robin boundary condition. In the opposite reflecting limit $q \to 0$ (Neumann boundary condition), the surface reaction is switched off, i.e., $\phi \to 0$, $p\expect{T} \to 1$, and $T = \delta \sim \mrm{Exp}(p)$. 
However, the acquired BLT $L = \ell_\delta$ does not vanish,                                               
\begin{equation}                                                                      
\label{eq:Llimit}                                                                     
\lim_{q \to 0} \expect{L}(\x_0) = \lim_{q \to 0} \frac{\phi(p,q|\x_0)}{q} = \frac{\partial \phi(p,q|\x_0)}{\partial q}\bigg|_{q=0} ,               
\end{equation}
which recovers the classical results for the BLT distribution for pure bulk decay \cite{Ye25JCP}.

\subsection{Universal law and generating function}
\label{sec:joint_gf}

Define the dimensionless \emph{cumulative risk} process:
\begin{equation}
\mathcal{R}_t = p t + q \ell_t.
\label{eq:cumrisk}
\end{equation}
Conditioned on the full trajectory $\X_t$, the thresholds are independent, so
for any $t>0$,
\begin{equation}
\p( T>t\,|\,\X_t) = \p\bigl(t <\delta \bigr) \p\bigl(\ell_t<\hat\ell \bigr) 
= e^{-p t} e^{ - q \ell_t} =e^{-\mathcal{R}_t}.
\end{equation}
The process $\mathcal{R}_t$ is continuous and strictly increasing from 0 to infinity for every trajectory. Note that the strict increase follows from $p>0$, while the boundary local time $\ell_t$ alone has plateaux. 
For each $z>0$, there exists a unique $t_z$ such that $\mathcal{R}_{t_z}=z$. Consequently, independent of the trajectory, the equality holds
\begin{equation}
\p(\mathcal{R}_{T}>z | \X_t ) = \p( T > t_z | \X_t ) = e^{-\mathcal{R}_{t_z}}=e^{-z}.
\end{equation}
As $e^{-z}$ is independent of the trajectory, the equality holds for arbitrary $\X_t$. Averaging over the reflected path yields 
$\p(\mathcal{R}_T>z)=e^{-z}$. 
Therefore, we obtain a universal law 
\begin{equation}
\mathcal{R}_{T} = p T + q L \sim \mrm{Exp}(1) ,
\label{eq:universal_exp}
\end{equation}
for any diffusive trajectory inside an arbitrary domain with $p,q>0$. 
The argument does not require a bounded domain: for $p>0$, the bulk decay guarantees $T \leqslant \delta < \infty$ and $\mathcal{R}_t \geqslant pt$ tends to infinity. 
Indeed, Eq.~\eqref{eq:universal_exp} is the cumulative-hazard time-rescaling principle realized by the boundary local time \cite{crowder2001classical, brown2002time}: $\mathcal{R}_t$ combines the ordinary time $t$ with the singular additive functional $\ell_t$. 
The exponential risk law is not specific to Brownian diffusion, but also valid for an arbitrary underlying motion across general classes of non-Markovian or anomalous transport \cite{bressloff2023encounter1}, whenever thresholds $\delta$ and $\hat{\ell}$ are independent of the underlying trajectory, and the total cumulative risk $\mathcal{R}_t$ is continuous, nondecreasing and diverges as $t$ tends to infinity. 
In the limit $p=0$, this condition can fail, e.g., for reflected Brownian motion outside a sphere in three dimensions.
%For unbounded domains in the limit $p=0$, stopping may fail with a positive probability, e.g., three-dimensional Brownian motion outside a sphere.  
The formula~\eqref{eq:universal_exp} can also be generalized to multiple clocks, for example, different bulk decay rates $p_i$ in disjoint subdomains and different surface reactivities $q_j$ on separated reactive regions (see Appendix~\ref{app:cumulative_risk} for details).

Since $\mathcal{R}_T$ has unit mean, we immediately recover the identity~\eqref{eq:sum_rule}. 
Moreover, all higher moments of the total risk are those of $\mrm{Exp}(1)$:
\begin{equation}
\E\!\left[(pT+qL)^n\right]
=\sum_{k=0}^n\binom{n}{k}p^k q^{n-k}\langle T^k L^{n-k}\rangle=n!.
\label{eq:all_moments}
\end{equation}
However, one cannot extract each individual term $\langle T^k L^{n-k} \rangle$ from Eq.~\eqref{eq:all_moments}. 
The two marginal quantities $S_T(t|\x_0)$ and $Q_L(\ell|\x_0)$ do not determine the joint distribution of $(T,L)$.
%The joint generating function naturally encompasses the whole joint distribution, which can be regarded as the two-dimensional Laplace transform. 
Let $h(t,\ell|\x_0)$ denote the generalized joint probability measure
of $(T,L)$ for the starting point $\x_0 \in \Omega$ under the joint stopping condition, including its singular contribution at $\ell=0$. To systematically access all joint moments of $T$ and $L$, we introduce the joint Laplace transform
\begin{align}
\label{eq:W_def}
W(\gamma,\eta|\x_0) &= \langle e^{-\gamma T-\eta L} \rangle(\x_0) \nonumber \\
&= \int_0^\infty \int_0^\infty \md t \, \md \ell \, h(t,\ell|\x_0) \, e^{-\gamma t-\eta\ell} ,
\end{align}
where $\gamma,\eta\geqslant 0$ and the average is over both the diffusion process and the independent exponential thresholds $\delta$ and $\hat\ell$. The stopping mechanism implies a natural decomposition of the random variable
$e^{-\gamma T-\eta L}$ according to which clock fires first:
\begin{equation}
\label{eq:exp_decomp}
e^{-\gamma T-\eta L}
= e^{-\gamma\delta-\eta\ell_\delta}\,\mathbf 1_{\delta<\tau}
+ e^{-\gamma\tau-\eta\hat\ell}\,\mathbf 1_{\tau<\delta} .
\end{equation}
Recall the splitting probability associated with the two stopping mechanisms
\begin{subequations}
\begin{align}
\phi(p,q|\x_0) &= \E_{\x_0} \left[ e^{-p\tau} \right] = \p_{\x_0}(\tau < \delta) , \\
1-\phi(p,q|\x_0) &= \E_{\x_0} \left[ e^{-q\ell_\delta} \right] = \p_{\x_0}(\ell_\delta < \hat\ell) .
\end{align}
\end{subequations}
Averaging $e^{-\gamma \delta-\eta\ell_\delta}1_{\delta<\tau}$ over $\delta$ conditioned on the trajectory gives $\int_0^\tau \md s \, p \, e^{-(p+\gamma)s} e^{-\eta \ell_s}$. 
Averaging the indicator $1_{s<\tau}$ over the threshold $\hat{\ell}$ yields $e^{-q \ell_s}$. The remaining path average is the shifted splitting probability, then averaging over $\X_t$ gives two transforms: 
\begin{subequations}
\begin{align}
\label{eq:bulk_term} 
W_\mrm{b}(\gamma,\eta|\x_0) &= \E_{\x_0} \left[e^{-\gamma\delta-\eta\ell_\delta}\mathbf 1_{\delta<\tau}\right] \nonumber \\ &= \frac{p}{p+\gamma}\,\bigl[1-\phi(p+\gamma,q+\eta|\x_0)\bigr], \\
\label{eq:surf_term}
W_\mrm{s}(\gamma,\eta|\x_0) &= \E_{\x_0} \left[e^{-\gamma\tau-\eta\hat\ell}\mathbf 1_{\tau<\delta}\right] \nonumber \\ &= \frac{q}{q+\eta}\,\phi(p+\gamma,q+\eta|\x_0) \, . 
\end{align}
\end{subequations}
Adding Eqs.~\eqref{eq:bulk_term} and \eqref{eq:surf_term}, the joint generating function takes the compact form
\begin{align}
\label{eq:W_alt}
W(\gamma,\eta|\x_0) &= W_\mrm{b}(\gamma,\eta|\x_0) + W_\mrm{s}(\gamma,\eta|\x_0) 
\\
&= \frac{p}{p+\gamma} + \frac{q\gamma-p\eta}{(p+\gamma)(q+\eta)} \phi(p+\gamma,q+\eta|\x_0) . \nonumber 
\end{align}
The differentiation of $W(\gamma,\eta|\x_0)$ at $(\gamma,\eta)=(0,0)$ gives
\begin{equation}
\expect{T^m L^n} (\x_0) = \left. (-1)^{m+n} \frac{\partial^{m+n} W(\gamma,\eta|\x_0)}{\partial \gamma^m \partial \eta^n} \right|_{\gamma=0, \eta=0} \, .
\label{eq:anymoment}
\end{equation}
See Appendix~\ref{app:Wrecur} for the moment hierarchy.

\subsection{Spectral duality}
\label{sec:dual}

The previous argument reduces the joint transform $W(\gamma,\eta|\x_0)$ to the function $\phi(p+\gamma,q+\eta|\x_0)$. 
Then we show how to compute the function $\phi(p,q|\x_0)$ through the Robin-Laplacian and generalized Steklov spectra. 

The survival probability satisfies $\partial_t S_\tau = D \Delta S_\tau$ in $\Omega$, and $-\partial_n S_\tau = q S_\tau$ on $\Gamma$. Consequently, the survival probability for the competing stopping problem satisfies 
\begin{subequations}
\label{eq:survival_pde}
\begin{align}
\partial_t S_T &= D\Delta S_T - p \, S_T \quad \text{in } \Omega, 
\label{eq:survival_pde_a} \\
-\partial_n S_T &= q \, S_T \qquad \qquad \quad \text{on } \Gamma,
\label{eq:survival_pde_b}
\end{align}
\end{subequations}                                                        
with initial condition $S_T(0|\x_0) = 1$.

For a bounded domain $\Omega$ with the Robin boundary condition on its reactive boundary $\Gamma$, the survival probability $S_\tau(t|\x_0)$ admits a spectral decomposition over the Laplacian eigenvalues $\lambda_n^{(q)}$ and eigenfunctions $u_n^{(q)}$: 
\begin{equation}
\label{eq:spectral_S_tau}
S_\tau(t|\x_0) = \sum_{n=0}^{\infty} a_n^{(q)} \, u_n^{(q)}(\x_0) \, e^{-D \lambda_n^{(q)} t} ,
\end{equation}
where $a_n^{(q)} = \int_{\Omega} \md \x \, u_n^{(q)}(\x)$ and eigenmodes $\{\lambda_n^{(q)}, u_n^{(q)}\}$ satisfy: 
\begin{subequations}
\label{eq:eigenvalue_problem}
\begin{align}
-\Delta u_n^{(q)} &= \lambda_n^{(q)} u_n^{(q)} \quad \text{in } \Omega, 
\label{eq:eigen_a} \\
- \partial_n u_n^{(q)} &= q \, u_n^{(q)} \qquad \text{on } \Gamma.
\label{eq:eigen_b}
\end{align}
\end{subequations}
Eigenvalues $\{\lambda_n^{(q)}\}$ form an increasing sequence, i.e., $0~<~\lambda_0^{(q)} \leq \lambda_1^{(q)} \leq \cdots \to \infty$. $\{u_n^{(q)}\}$ are orthonormal in $L^2(\Omega)$, i.e., $\int_\Omega \md \x \, u_n^{(q)}(\x) u_m^{(q)}(\x) = \delta_{nm}$. 
As a consequence, the spectral decomposition of $S_T(t|\x_0)$ reads:
\begin{equation}                                                                      
\label{eq:spectral_S}                                                                 
S_T(t|\x_0) = \sum_{n=0}^{\infty} a_n^{(q)} \, u_n^{(q)}(\x_0) \, e^{-(D\lambda_n^{(q)} + p)t} . 
\end{equation}
The spatial eigenfunctions $\{u_n\}$ are the same as in Eq.~\eqref{eq:eigen_a}, since the constant decay operator $p$ commutes with the Laplacian. 
However, each eigenmode decays with an enhanced rate $D\lambda_n^{(q)} + p$. 
Equivalently, the operator $\mathcal{L}_p = -(\Delta - p/D)$ in Eq.~\eqref{eq:survival_pde_a} corresponds to eigenvalues shifted as $\lambda_n^{(q)} \to \lambda_n^{(q)} + p/D$.

Via Eqs. \eqref{eq:H_def} and \eqref{eq:spectral_S_tau}, the spectral representation of $\phi(p,q|\x_0)$ under the eigenpairs $\{\lambda_n^{(q)}, u_n^{(q)}\}$ reads 
\begin{equation}                                                                      
\label{eq:H_spectral}                                                                 
\phi(p,q|\x_0) = \sum_{n=0}^{\infty} a_n^{(q)} \, u_n^{(q)}(\x_0)\, \frac{D\lambda_n^{(q)}}{D\lambda_n^{(q)} + p}, 
\end{equation}                                                                        
showing that the bulk decay rate $p$ acts as a cutoff that suppresses the contribution of low-frequency modes. More precisely, modes with $D\lambda_n^{(q)} \gg p$ contribute the weight $a_n^{(q)} u_n^{(q)}(\x_0)$, whereas modes with $D\lambda_n^{(q)} \ll p$ are suppressed by the factor $D\lambda_n^{(q)}/p$. 
As a result, we have the spectral representation of $\expect{T}(\x_0)$: 
\begin{equation} 
\expect{T}(\x_0) = \sum_{n=0}^{\infty} \frac{a_n^{(q)} \, u_n^{(q)}(\x_0)}{D\lambda_n^{(q)} + p} . 
\end{equation}
Recall that $\sum_n a_n^{(q)} \, u_n^{(q)}(\x_0) = 1$.

On the reactive boundary $\Gamma$, $Q_\delta(\ell|\x_0)$ evolves with boundary local time according to the Dirichlet-to-Neumann operator $\mathcal{M}_p$: 
$\partial_\ell Q_\delta=-\mathcal{M}_p Q_\delta$. The operator $\mathcal{M}_p$ maps $H^{1/2}(\Gamma) \to H^{-1/2}(\Gamma)$, and $\mathcal{M}_p w = \partial_n s |_\Gamma$, where $s$ satisfies $(p-D\Delta) s = 0$ in $\Omega$ and $s = w$ on $\Gamma$. 
Hence we obtain
\begin{equation}
\label{eq:DtNQL}
\partial_\ell Q_L = - \mathcal{M}_p Q_L - q \, Q_L . 
\end{equation}

The distribution of $\ell_\delta$, evaluated at an independent exponential lifetime $\delta$, is known explicitly in terms of the generalized Steklov spectrum \cite{Grebenkov2019, Ye25JCP}. 
We can consider the statistics of $\ell_\delta$ directly in the Laplace domain, which satisfies the spectral expansion 
\begin{align}
\Psi(\ell,p|\x_0) &= p \, \tilde{S}_\infty (p|\x_0) \delta_0(\ell) \nonumber \\
&\quad + \sum_{k=0}^\infty b_k^{(p)} [V_k^{(p)}(\x_0)]^\ast \mu_k^{(p)} e^{-\mu_k^{(p)} \ell} , 
\label{eq:Psibk}
\end{align}
where 
$\tilde{S}_\infty (p|\x_0)$ is the Laplace-transformed survival probability in the presence of a perfectly absorbing boundary, 
$b_k^{(p)} = \int_\Gamma \md S \, V_k^{(p)}$, and 
$\mu_k^{(p)}$, $V_k^{(p)}(\x)$ are respectively the eigenvalues and eigenfunctions of the generalized Steklov problem:  
\begin{subequations}
\label{eq:Steklov}
\begin{align}
(p-D\Delta) V_k^{(p)} &= 0 \qquad \qquad \text{in } \Omega , \\
\partial_n V_k^{(p)} &= \mu_k^{(p)} V_k^{(p)} \quad \text{on } \Gamma .
\end{align}
\end{subequations}
Compared with Eq.~\eqref{eq:BLTPsi}, the equality holds $\pi_0(p|\x_0) = p \tilde{S}_\infty (p|\x_0)$. 
The spectrum of the generalized Steklov problem~\eqref{eq:Steklov} is known to be discrete \cite{Levitin2023}. 
The eigenpairs $\{ \mu_k^{(p)}, V_k^{(p)} \}$ are enumerated by $k=0,1,2,\ldots$ to order the eigenvalues into an increasing sequence, and 
eigenfunctions satisfy $\int_\Gamma \md S \, [V_j^{(p)} ]^\ast V_k^{(p)}  = \delta_{jk}$. 
Furthermore, the generalized Steklov spectrum gives 
\begin{subequations}
\begin{align}
Q_\delta(\ell|\x_0) &= \sum_{k=0}^\infty b_k^{(p)} [V_k^{(p)}(\x_0)]^\ast e^{-\mu_k^{(p)} \ell}, \label{eq:Qdspec} \\
Q_L(\ell|\x_0) &= \sum_{k=0}^\infty b_k^{(p)} [V_k^{(p)}(\x_0)]^\ast e^{- (\mu_k^{(p)}+q) \ell}. 
\end{align}
\end{subequations}
Similarly, from the generalized Steklov problem, the spectrum of $-(\mathcal{M}_p + q)$ in Eq.~\eqref{eq:DtNQL} shifts eigenvalues from $-\mu_k^{(p)}$ to $-(\mu_k^{(p)} + q)$. 
Moreover, let $U(\ell,t|\x_0)$ denote the density of the first-crossing time $\mathscr{T}_\ell$ \cite{Ye26JCP}, its Laplace transform 
$\tilde{U}(\ell,p|\x_0)= \int_0^\infty \md t \, e^{-pt} \, U(\ell,t|\x_0)$ admits the spectral expansion \cite{Grebenkov2019},
\begin{equation}
\tilde{U}(\ell,p|\x_0) =  \sum_{k=0}^\infty e^{-\mu_k^{(p)} \ell} b_k^{(p)} [V_k^{(p)}(\x_0)]^\ast . 
\label{eq:Utilde}
\end{equation}
Indeed, this coincides with $Q_\delta(\ell)$ shown in Eq.~\eqref{eq:Qdspec}. 
The pure surface-reaction survival probability $S_\tau(t|\x_0)$ can be expressed in terms of $U(\ell,t|\x_0)$ as 
\begin{equation}
S_\tau(t|\x_0) = \int_t^\infty \md t^\prime \, \int_0^\infty \md \ell \, q \, e^{-q\ell} U(\ell,t^\prime|\x_0) , 
\end{equation} 
i.e., survival up to $t$ requires the crossing of $\hat{\ell}$ to occur later. 
As a consequence, from Eqs.~\eqref{eq:H_def} and \eqref{eq:Utilde}, the spectral representation of $\phi(p,q|\x_0)$ under the eigenpairs $\{ \mu_k^{(p)}, V_k^{(p)} \}$ reads 
\begin{align}
\label{eq:H_spectral2}
\phi(p,q|\x_0) &= q\int_0^\infty \md t\, e^{-pt} \int_0^\infty \md\ell \, e^{-q\ell} U(\ell,t|\x_0) \nonumber \\
&= \sum_{k=0}^\infty b_k^{(p)} [V_k^{(p)}(\x_0)]^\ast \frac{q}{\mu_k^{(p)} + q} . 
\end{align}
Eigenmodes with $\mu_k^{(p)} \gg q$ are suppressed by the factor $q/\mu_k^{(p)}$, 
whereas eigenmodes with $\mu_k^{(p)} \ll q$ contribute the weight $b_k^{(p)} [V_k^{(p)}(\x_0)]^\ast$. 
As a consequence, we obtain the spectral expansion of mean boundary local time:
\begin{equation}
\expect{L}(\x_0) = \sum_{k=0}^\infty \frac{b_k^{(p)} [V_k^{(p)}(\x_0)]^\ast}{\mu_k^{(p)} + q} . 
\end{equation}

From Eqs. \eqref{eq:H_spectral} and \eqref{eq:H_spectral2}, 
since both spectral series represent the same probability $\phi(p,q|\x_0)$, they are identically equal for all $\x_0$, $p>0$, and $q>0$. This yields the spectral dual identity: 
\begin{equation}
\sum_{n=0}^{\infty} \frac{D\lambda_n^{(q)} \, a_n^{(q)} \, u_n^{(q)}(\x_0)}{D\lambda_n^{(q)} + p} = \sum_{k=0}^\infty \frac{q \, b_k^{(p)} [V_k^{(p)}(\x_0)]^\ast}{\mu_k^{(p)} + q} ,
\label{eq:duality}
\end{equation}
which reflects the equivalence of describing the competition between bulk decay and surface reaction either in the time domain (Robin-Laplacian spectrum) or in the boundary local time domain (generalized Steklov spectrum). 
In practice, we may choose whichever side is more convenient for calculations.
For instance, if one needs derivatives $\partial_p \phi$ or $\partial_q \phi$, the former can be obtained  from the Robin-Laplacian spectrum as $q$ is fixed, while the latter can be obtained from the generalized Steklov spectrum as $p$ is fixed: 
\begin{subequations}
\label{eq:phi_derivatives}
\begin{align}
\phi_p \equiv \partial_p \phi &= -\sum_n a_n^{(q)} u_n^{(q)}(\x_0) \frac{D\lambda_n^{(q)}}{(p+D\lambda_n^{(q)})^2} ,
\label{eq:phi_p} \\
\phi_q \equiv \partial_q \phi &= \sum_k b_k^{(p)} [V_k^{(p)}(\x_0)]^\ast \frac{\mu_k^{(p)}}{(\mu_k^{(p)}+q)^2} .
\label{eq:phi_q}
\end{align}
\end{subequations}
These results are consistent with the physical intuition that increasing $q$ favors the surface reaction and increases $\phi$, whereas increasing $p$ favors the bulk decay and decreases $\phi$.

\section{Explicit solutions in a three-dimensional ball}
\label{sec:example}

The probabilistic and spectral constructions are valid for an arbitrary smooth confining domain, e.g., droplets, vesicles, or spherical microreactors.  
In this section, to illustrate these results explicitly, we consider the three-dimensional ball $\Omega = \{|\x| < R\} \subset \mathbb{R}^3$ with a homogeneous partially reactive boundary $\partial\Omega = \Gamma$, where rotational symmetry implies that all quantities depend only on $r = |\x_0|$.  
Hence only the spherically symmetric Steklov eigenmode contributes to the expansions in Eq.~\eqref{eq:H_spectral2}:
\begin{subequations}
\label{eq:Steklov0}
\begin{align}
V_0^{(p)}(r) &= \frac{1}{\sqrt{|\Gamma|}} \frac{i_0(\alpha r)}{i_0(\alpha R)} , 
\label{eq:V0} \\
\mu_0^{(p)} &= \alpha \coth(\alpha R) - \frac{1}{R} ,
\label{eq:mu0}
\end{align}
\end{subequations}
where $\alpha = \sqrt{p/D}$ and $i_0(z) = \sinh(z)/z$ is the modified spherical Bessel function of the first kind.     
From Eq.~\eqref{eq:H_spectral2} we obtain the splitting probability $\phi(p,q|r)$: 
\begin{align}                                                                      
\label{eq:H_ball}                                                                     
\phi(p,q|r) &= \frac{q \, b_0^{(p)} \, V_0^{(p)}(r)}{\mu_0^{(p)} + q} \nonumber \\
&= \frac{q R^2 \sinh \left(\alpha r \right)}{r \left[ (q R-1) \sinh \left(\alpha R\right)+ \alpha R \cosh \left(\alpha R\right) \right]} , 
\end{align}
where $b_0^{(p)} = \sqrt{|\Gamma|}$. 
Once $\phi(p,q|r)$ is known, all moments follow from the formulas \eqref{eq:anymoment}. 
For other domains (interval, disk, circular annulus, and spherical shell), see Appendix~\ref{app:moredomains} for details.

\subsection{Fixed starting positions and limiting regimes}
\label{app:asymptotics}

In the limit $q \to 0$ with fixed $p$, the surface reaction is rare and %$\phi=O(q)$: 
\begin{equation}                                                                      
\phi(p,q|r) \approx \frac{q R^2 \sinh \left(\alpha r \right)}{r \left[ \alpha R \cosh \left(\alpha R\right)- \sinh \left(\alpha R\right) \right] }. 
\end{equation} 
In this limit, the particle almost always decays in the bulk. 
The acquired boundary local time $L$ is mostly the local time accumulated up to the random lifetime $\delta$, and its mean value is independent of $q$: 
\begin{equation}                                                                      
\expect{L}(p,q|r) \approx \frac{R^2 \sinh \left(\alpha r \right)}{r \left[ \alpha R \cosh \left(\alpha R\right)- \sinh \left(\alpha R\right) \right] }.  
\end{equation} 
In the limit $q \to \infty$ with fixed $p$, we obtain 
\begin{equation}
\phi(p,q|r) \approx \frac{R \sinh (\alpha  r) }{r \sinh (\alpha  R)}  + \frac{\sinh (\alpha  r) [1-\alpha  R \coth (\alpha  R)]}{q \, r \sinh (\alpha  R)} . 
\end{equation}
In this Dirichlet limit, every particle that reaches the boundary reacts immediately.  
In particular, in the limit $p\to0$, we recover the mean first-passage time:
\begin{equation}
\label{eq:FPTlimite}
\expect{T}(r) = \lim_{p\to0} \frac{1}{p} \left[ 1-  \frac{R \sinh (\alpha  r) }{r \sinh (\alpha  R)} \right] = \frac{R^2-r^2}{6 D} . 
\end{equation}

In turn, in the limit $p \to 0$ with fixed $q$, we obtain 
\begin{equation}                                                                      
\label{eq:MFRT}
\phi(p,q|r) \approx 1-p \left( \frac{R^2 - r^2}{6 D} + \frac{R}{3Dq} \right) . 
\end{equation}                                                                        
The coefficient of $-p$ is precisely the mean surface-reaction time in the absence of bulk decay. Its first term is the mean time to reach the boundary (Eq.~\eqref{eq:FPTlimite}), whereas the second term accounts for the additional delay caused by a finite surface reactivity $q$. 
In the limit $p \to \infty$ with fixed $q$, $\phi(p,q|r)$ is exponentially small for $r \in (0,R)$ and asymptotically equals 
\begin{equation}
\label{eq:phipapp}
\phi(p,q|r) \approx \frac{qR^2 e^{-\alpha(R-r)}}{r(\alpha R + qR - 1)} . 
\end{equation}

\begin{figure}[t!]                                                                  
\includegraphics[width=\linewidth]{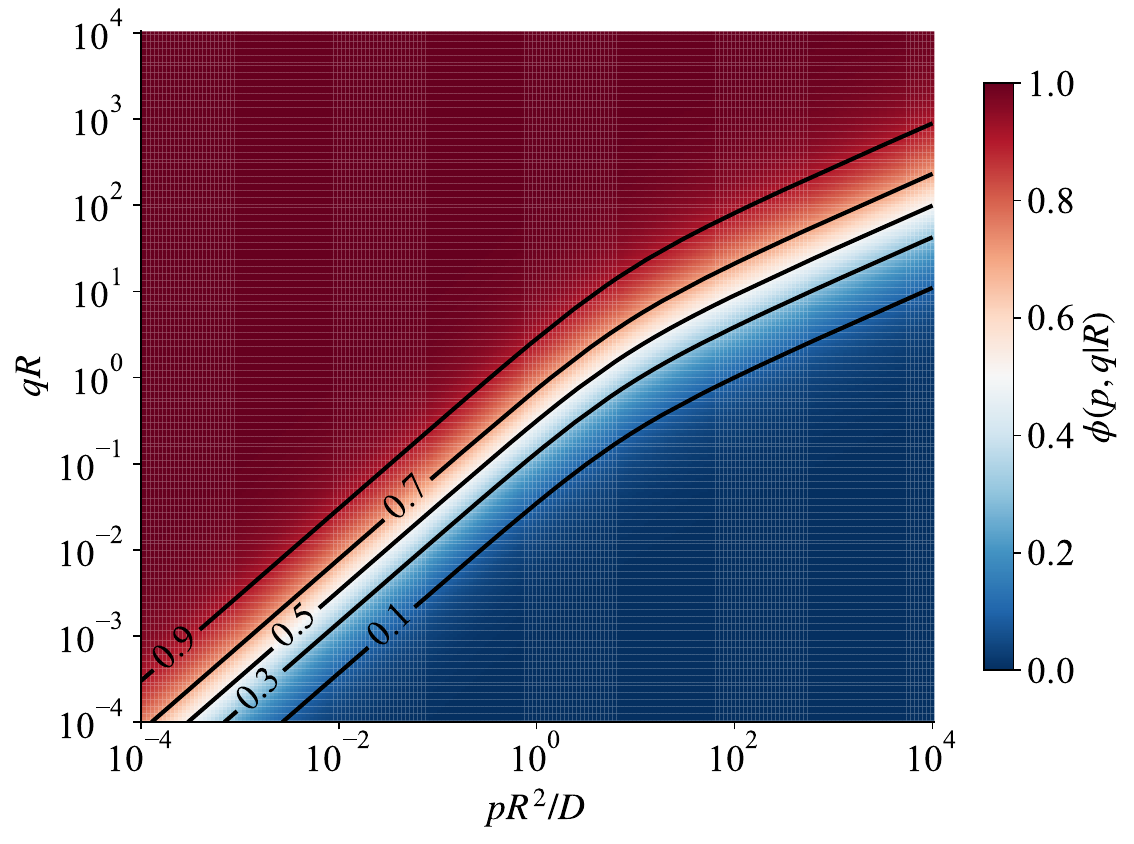} 
\caption{Heat map of the splitting probability $\phi(p,q|R)$ for starting points on the reactive boundary of a three-dimensional ball ($R=1$, $D=1$) over the $(pR^2/D,qR)$ plane in log-log scale. 
Black curves indicate contour lines at $\phi_c = 0.1$, $0.3$, $0.5$, $0.7$, $0.9$.
\label{fig:corrR_app}}
\end{figure}

Physically, the radial dependence of $\phi(p,q|r)$ reflects the competition between bulk decay and diffusion to the boundary. For a particle starting near the center $r \to 0$, the rotational symmetry removes the linear dependence on $r$, since $i_0(\alpha r)=1+O(r^2)$, and the bulk decay can occur before the particle reaches the surface. 
When particles are initially on the reactive surface $\Gamma$, i.e., $r=R$, only the rotationally invariant Steklov eigenvalue contributes to $\phi(p,q|R)$: 
\begin{equation}
\phi(p,q|R) = \frac{q R}{\alpha R \coth \left(\alpha R\right)+q R-1} = \frac{q}{\mu_0^{(p)} + q} . 
\label{eq:phiR}
\end{equation}
Presented in Fig. \ref{fig:corrR_app}, the heat map of Eq.~\eqref{eq:phiR} is divided into three regimes: 
in the lifetime-dominated regime shown in blue ($\phi < 0.1$), most particles decay before undergoing a surface reaction, so $p\expect{T}\approx1$ and $q\expect{L}\ll1$; 
in the surface-dominated regime shown in red ($\phi > 0.9$), the surface reaction occurs first for most trajectories, so $q\expect{L}\approx1$ and $p\expect{T}\ll1$; 
and the mixed regime ($0.1 < \phi < 0.9$) is where both mechanisms contribute and where the competition is most interesting. 
Contour lines $\phi_c$ exhibit two distinct slopes in log-log scale: 
for $pR^2/D \ll 1$, $\alpha R \coth \left(\alpha R\right) \approx 1 + pR^2/(3D)$, then
\begin{equation}
qR \approx \frac{\phi_c}{1-\phi_c} \frac{pR^2}{3D} ,
\end{equation}
whereas for $pR^2/D \gg 1$, $\coth \left(\alpha R\right) \approx 1$, thus
\begin{equation}
qR \approx \frac{\phi_c}{1-\phi_c} \sqrt{\frac{pR^2}{D}} . 
\end{equation}
See the heat map of $\expect{T}(p,q|R)$ and $\expect{L}(p,q|R)$ in Appendix~\ref{app:H_derivatives_ball}.

\subsection{Uniformly distributed starting points in the ball} 
\label{sec:phase}

\begin{figure}[t!]                                                                  
\includegraphics[width=\linewidth]{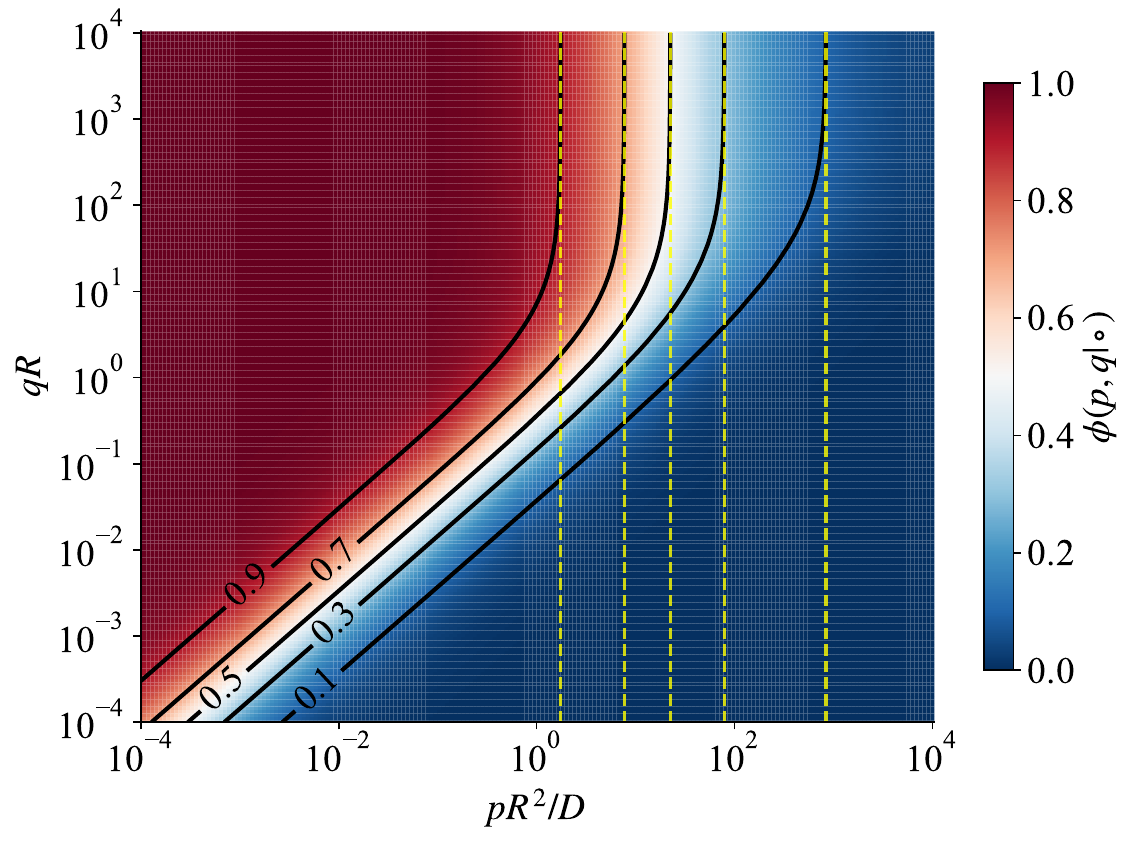} 
\caption{
Heat map of the splitting probability $\phi(p,q|\circ)$ for uniformly distributed starting points in the three-dimensional ball ($R=1$, $D=1$) over the $(pR^2/D,qR)$ plane in log-log scale. 
Black curves indicate contour lines at $\phi_c = 0.1$, $0.3$, $0.5$, $0.7$, $0.9$.
Yellow dashed lines mark the corresponding crossover values $p_c$ solved from $\phi_c = \phi_\infty(p_c|\circ)$ (Eq.~\eqref{eq:phiasymp}). 
Each $(\phi_c, p_c)$ reads $(0.1,838.927)$, $(0.3,78.7298)$, $(0.5,22.4043)$, $(0.7,7.69806)$, and $(0.9,1.74768)$. 
\label{fig:phiunif}} 
\end{figure}

To probe the global effect of confinement, we next average over uniformly distributed starting points in the ball of radius $R$. The splitting probability reads 
\begin{align}
\label{eq:phiunif}
\phi(p,q|\circ) &= \int_0^R \md r \, \frac{3r^2}{R^3} \, \phi(p,q|r) \\
&= \frac{3 D q \left[ \alpha R \cosh \left( \alpha R\right)-\sinh \left(\alpha R\right)\right]}{p R\left[(q R-1) \sinh \left(\alpha R\right)+ \alpha R \cosh \left(\alpha R\right)\right]} , \nonumber 
\end{align}
where $\circ$ indicates the uniform spatial distribution. 
Figure~\ref{fig:phiunif} displays Eq.~\eqref{eq:phiunif} as the heat map over the $(pR^2/D, qR)$ plane in log-log scale. 
Black curves are contour lines for $\phi(p,q|\circ) = \phi_c$ and can be parameterized by either $p$ or $q$.

Unlike Fig. \ref{fig:corrR_app}, 
Fig. \ref{fig:phiunif} exhibits almost vertical contour lines in the limit $q\to\infty$, which indicates a crossover value $p_c$. 
In other words, even a perfectly absorbing boundary cannot compensate for a sufficiently rapid bulk decay when the starting point is uniformly distributed in the volume. 
Taking the limit $q\to\infty$ of Eq.~\eqref{eq:phiunif}, we denote 
\begin{equation}
\label{eq:phiasymp}
\phi_\infty(p|\circ) = \lim_{q\to\infty} \phi(p,q|\circ) = \frac{3 \left[ \alpha R \coth \left(\alpha R\right)-1\right]}{\alpha^2 R^2} . 
\end{equation}
In this case, the limit $\phi_\infty(p_c)$ equals $\phi_c$ at $p=p_c$. 
Since $\phi_\infty(p)$ is a monotonically decreasing function of $p$, 
$\phi_\infty(p) < \phi_c$ if $p>p_c$, and vice versa.  
We can solve Eq.~\eqref{eq:phiasymp} numerically for the yellow asymptotic lines in Fig. \ref{fig:phiunif}.

\begin{figure}[t!]
    \includegraphics[width=\linewidth]{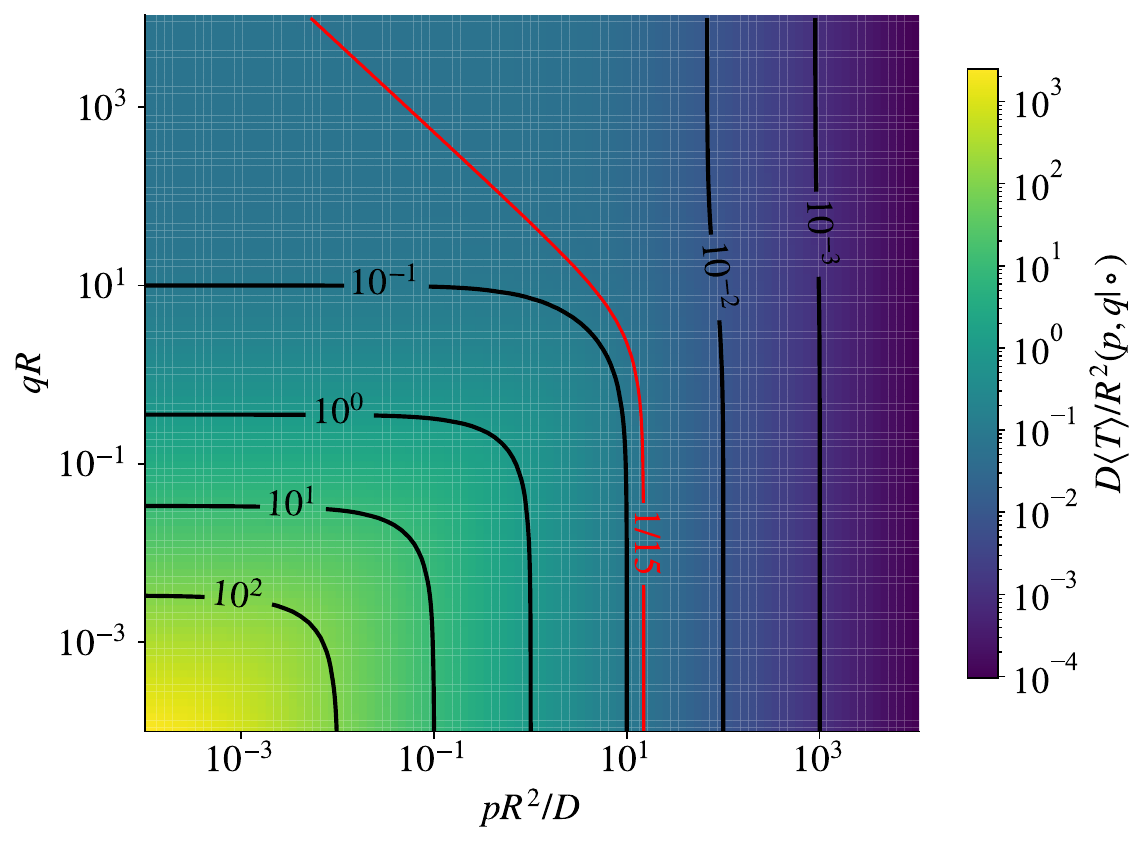}
    \rlap{\hspace{-0.5\linewidth}\raisebox{-0.0\linewidth}{\textbf{(a)}}}
\\
    \includegraphics[width=\linewidth]{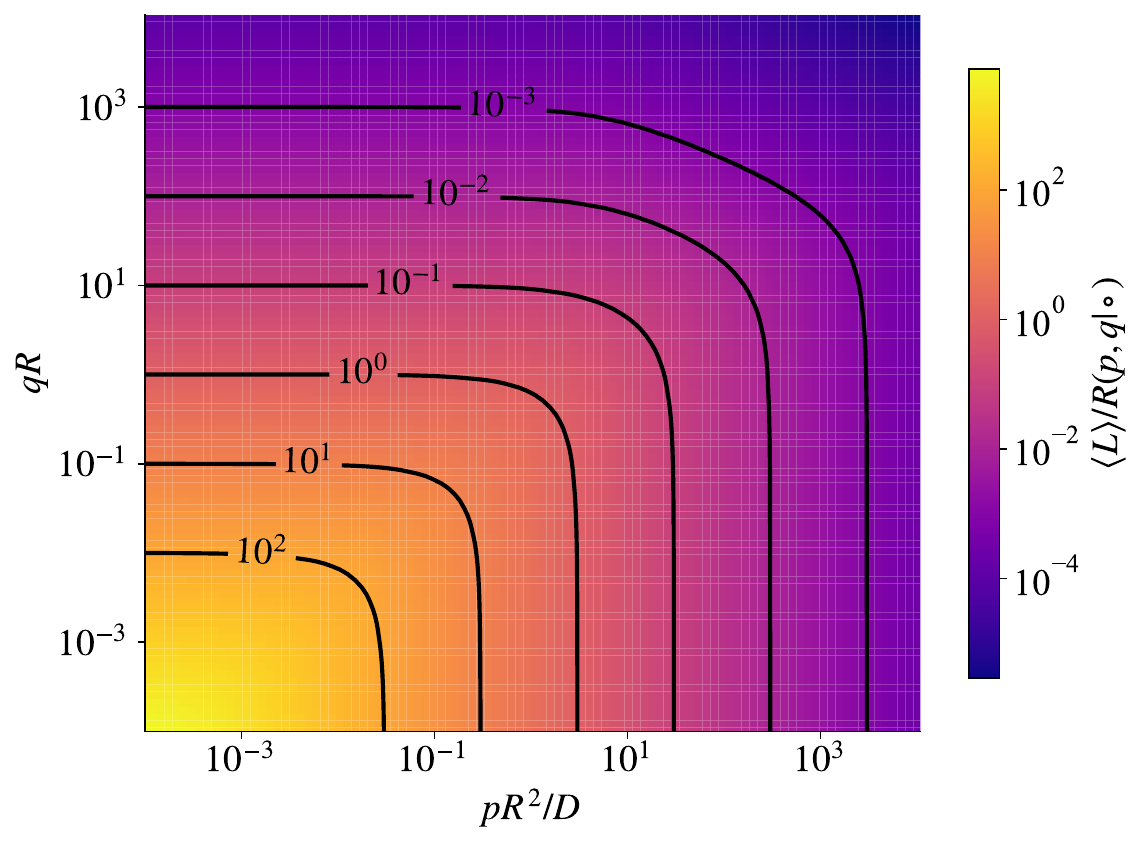}
    \rlap{\hspace{-0.5\linewidth}\raisebox{-0.0\linewidth}{\textbf{(b)}}}
\caption{Heat maps for uniformly distributed starting points in the ball ($R=1$, $D=1$) over the $(pR^2/D,qR)$ plane in log-log scale. 
\textbf{(a)} Mean stopping time $\expect{T}(p,q|\circ)$. 
Black curves indicate contour lines for $\expect{T}(p,q|\circ) = 10^{-3}$, $10^{-2}$, $10^{-1}$, $10^0$, $10^{1}$, $10^{2}$, while the red line is the contour line for $T_\mrm{D} = R^2/(15D) = 1/15$.  
\textbf{(b)} Mean acquired boundary local time $\expect{L}(p,q|\circ)$. 
Black curves indicate contour lines for $\expect{L}(p,q|\circ) = 10^{-3}$, $10^{-2}$, $10^{-1}$, $10^0$, $10^{1}$, $10^{2}$. 
}
\label{fig:meano}
\end{figure}

Based on Eq.~\eqref{eq:phiunif}, we can derive the mean stopping time $\expect{T}(p,q|\circ)$ and the mean acquired boundary local time $\expect{L}(p,q|\circ)$. 
\begin{subequations}
\begin{align}
\label{eq:meanTunif}
\expect{T}(p,q|\circ) &= \frac{1}{p}-\frac{3 D q}{p^2 R} \\
& \quad + \frac{3 D q^2 \sinh (\alpha  R)}{p^2 [ (q R-1) \sinh (\alpha  R)+\alpha  R \cosh (\alpha  R) ]} , \nonumber \\ 
\label{eq:meanLunif}
\expect{L}(p,q|\circ) &= \frac{3 D \left(\alpha  R \cosh \left(\alpha  R\right)-\sinh \left(\alpha  R\right)\right)}{p R \left[(q R-1) \sinh \left(\alpha  R\right)+\alpha  R \cosh \left(\alpha  R\right)\right]} . 
\end{align}
\end{subequations}
For uniformly distributed starting points in the ball of radius $R$, the mean first-passage time is $T_\mrm{D} = R^2/(15D)$, which corresponds to $p=0$ and $q=\infty$. 
If $q=\infty$ but $p>0$, a fraction of particles decay in the bulk without reaching the absorbing boundary, hence the mean stopping time is less than $T_\mrm{D}$. 
On the other hand, if $p=0$ but $q\in(0,\infty)$, we obtain the mean first-reaction time as $T_\mrm{D} + R/(3Dq)$. Thus any finite reactivity $q$ increases the mean stopping time. 
Eq.~\eqref{eq:meanTunif} is exhibited as the heat map in Fig.~\ref{fig:meano}(a) for the case $R=1$ and $D=1$, where 
$T_\mrm{D}$ in red contour line divides the heat map into two parts. 
On the left side $\expect{T}(p,q|\circ) > T_\mrm{D}$, contour lines approach finite $q$ as $p\to0$, whereas on the right side $\expect{T}(p,q|\circ) < T_\mrm{D}$, they diverge and there exists a minimal $p$ value for a given mean stopping time. In the limit $q\to\infty$, Eq.~\eqref{eq:meanTunif} reduces to: 
\begin{equation}
\expect{T}_\infty(p|\circ) = \lim_{q\to\infty} \expect{T}(p,q|\circ) = \frac{1}{p} + \frac{3 D}{p^2 R^2}-\frac{3 D \alpha \coth \left(\alpha R\right)}{p^2 R} . 
\label{eq:Tinfty}
\end{equation}
This quantity decreases as $p$ increases, and when $p\to0$, it reduces to $T_\mrm{D}$. 
For a prescribed mean below $T_\mrm{D}$, the right-hand contours terminate at a finite $p$ determined by Eq.~\eqref{eq:Tinfty} numerically. 
In addition, in Fig. \ref{fig:meano}(a), we observe that as $q \to 0$, $\phi \to 0$ and $\expect{T} \simeq 1/p$.

For $\expect{L}(p,q|\circ)$ presented in Fig.~\ref{fig:meano}(b), contour lines exhibit no contour-topology change analogous to that in $\expect{T}(p,q|\circ)$. 
The second moments $\expect{T^2}(\circ)$, $\expect{L^2}(\circ)$ are given in Appendix~\ref{app:H_derivatives_ball}, 
and the correlation coefficient $C(p,q|\circ)$ is derived with the heat map illustration.

\subsection{Numerical simulations}
\label{sec:numerical}                                   

Beyond the theoretical results, we perform Monte Carlo simulations for the joint stopping problem, in order to validate the universal cumulative-risk law,  
and test numerical algorithms that generalize to complex geometries where explicit solutions are unavailable.                         

\begin{figure}[t!]                                                                  
\includegraphics[width=\linewidth]{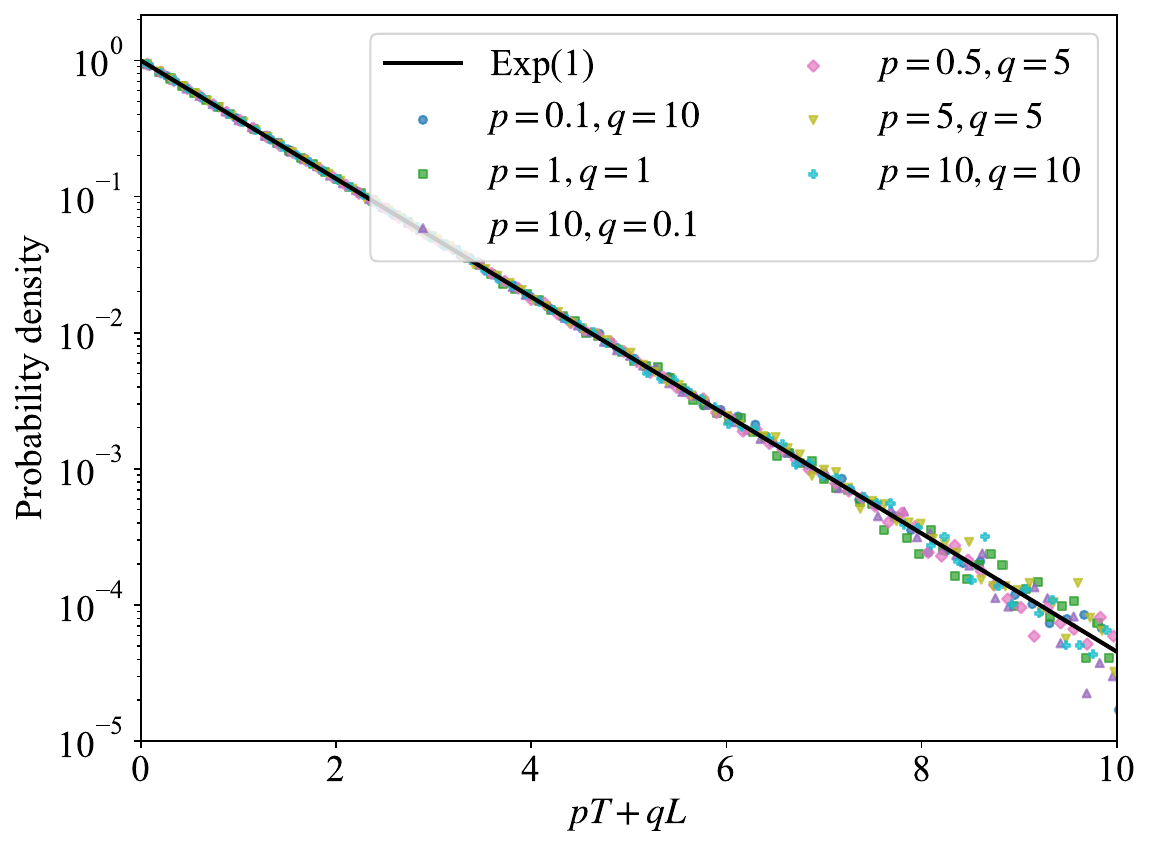}                                 
\caption{Empirical probability densities of the cumulative risk $\mathcal{R}_T = pT+qL$ obtained by Monte Carlo simulations for several $(p,q)$ rate pairs (symbols), compared with the unit exponential density (black line).
For each $(p,q)$ pair, starting points are uniformly distributed in the ball ($R=1$, $D=1$). 
Each dataset contains $N=10^6$ independent trajectories generated by the WoS-EFL algorithm with the boundary layer thickness $\varepsilon = 10^{-3}$. 
\label{fig:Expdist}} 
\end{figure}

For high accuracy and efficiency, especially near boundaries, the preferred method is the walk-on-spheres (WoS) algorithm \cite{Muller56, Binder12, Sawhney2020} combined with the escape-from-a-layer (EFL) method for BLT sampling \cite{Ye25JCP}. 
The WoS-EFL algorithm for the joint problem can be summarized as follows. 
\begin{enumerate}
%(i) 
\item For each independent trajectory $\X_t$, generate the uniformly distributed starting point $\X_{t=0} = \x_0$ inside the unit ball and generate two independent random thresholds: $\delta \sim \mrm{Exp}(p)$ and $\hat{\ell} \sim \mrm{Exp}(q)$. 
%(ii) 
\item At each numerical step, compute the distance $d = |\x_t - \Gamma|$ from the current position $\x_t$ to the reactive surface $\Gamma$. 
%(iii) 
\item Sample the exit position on the sphere of radius $d$ and the exiting time $\Delta t$ in the WoS framework \cite{grebenkov2014efficient}, update time $t$ by $\Delta t$. 
%(iv) 
\item If the cumulative time $t$ exceeds $\delta$, record $T=\delta$, $L=\ell_\delta$, and stop. 
%(v) 
\item If $d<\varepsilon$, where $\varepsilon$ is a prescribed layer thickness much smaller than the domain size, we
regard the particle as having entered the boundary layer and use the EFL approach \cite{Ye25JCP} to sample the escape position, the time increment $\Delta t$, and the boundary local time increment $\Delta\ell$, update the boundary local time $\ell_t$ with $\Delta\ell$. 
%(vi) 
\item If the accumulated boundary local time $\ell_t$ exceeds $\hat{\ell}$, record $L=\hat{\ell}$ and $T = \mathscr{T}_L$ as the interpolated hitting time, then stop. 
%(vii) 
\item Repeat the whole procedure for $N$ independent trajectories, and record their stopping time $T$, acquired boundary local time $L$, and stopping mechanism $M$ (bulk or surface). 
We define $M = 0$ for the bulk decay, while $M=1$ for the surface reaction. 
If one EFL step crosses both remaining thresholds, the mechanism with the smaller relative increment is selected. 
\end{enumerate}

The universal exponential law \eqref{eq:universal_exp} can be validated through the recorded numerical dataset above. 
Figure \ref{fig:Expdist} presents the probability density of $\mathcal{R}_T = pT+qL$ under different parameters in semi-log scale, where 
all markers collapse onto the unit exponential density. 
Agreement for $\mathcal{R}_T$ alone does not determine the joint law of $T$ and $L$. 
Moreover, we can also estimate the splitting probability $\phi$ and moments: 
\begin{subequations}
\begin{align}
\label{eq:phiest}
\phi_\mrm{est} &= \frac{1}{N} \sum_{i=1}^N M_i , \\
\label{eq:TLtest}
\expect{T^m L^n}_\mrm{est} &= \frac{1}{N} \sum_{i=1}^N (T_i)^m (L_i)^n  . 
\end{align}
\end{subequations}
See Appendix~\ref{app:otherphir} for more numerical results, including the splitting probability and the correlation coefficient.

\subsection{Comparison of spectral representations}

\begin{figure}[t!]
\includegraphics[width=\linewidth]{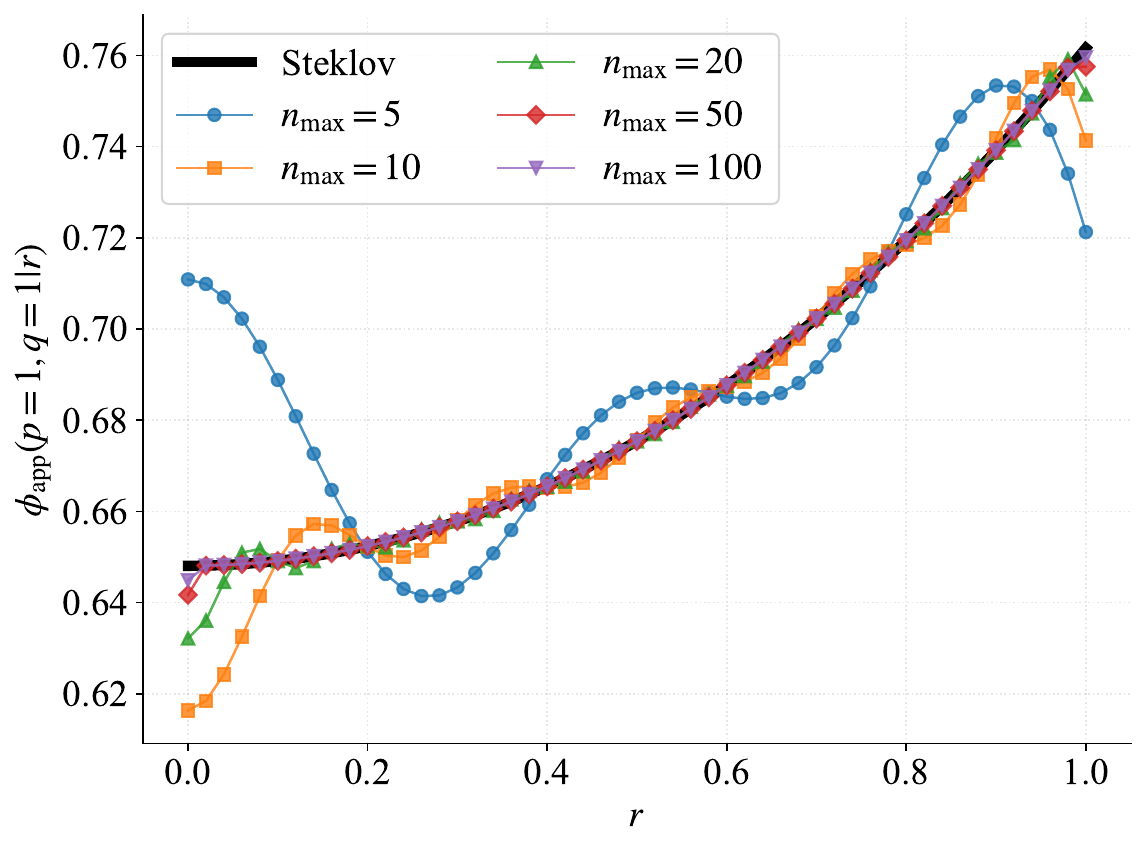} 
\caption{
Comparison of the exact generalized Steklov expression for $\phi(p=1,q=1|r)$ (Eq.~\eqref{eq:H_ball}, black curve)
with truncated Robin-Laplacian approximations (Eq.~\eqref{eq:phiappRobin}, colored curves with symbols) in the ball ($R=1$, $D=1$). 
The truncation orders are $n_\mrm{max}=5$ (circles), $10$ (squares), $20$ (upward triangles), $50$ (diamonds), and $100$ (downward triangles). 
\label{fig:duality}}
\end{figure}

In Sec. \ref{sec:dual}, $\phi(p,q|\x_0)$ is expressed by the spectral representations in the Robin-Laplacian spectrum (Eq.~\eqref{eq:H_spectral}) and the generalized Steklov spectrum (Eq.~\eqref{eq:H_spectral2}). 
For a three-dimensional ball, only the spherically symmetric Steklov mode contributes (Eq.~\eqref{eq:H_ball}). 
Furthermore, the spherically symmetric eigenmodes of Robin-Laplacian read
\begin{equation}
u_n^{(q)}(r) = A_n \frac{\sin(k_n r)}{k_n r} , \qquad \lambda_n^{(q)} = k_n^2 ,
\end{equation}
where $k_n = x_n/R$ ($n=0, 1, 2, \dots$), 
and $x_n$ is the $n$-th positive root of equation $x\cot x = 1-qR$. 
Since $\int_\Omega \md \x \, \left\vert u_n^{(q)}(\x) \right\vert^2=1$, the normalization constant  $A_n$ reads 
\begin{equation}
A_n = \frac{x_n}{\sqrt{2\pi R^3\left(1-\dfrac{\sin 2x_n}{2x_n}\right)}}.
\end{equation}
Equivalently, $u_n^{(q)}(r)$ can be written as
\begin{equation}
u_n^{(q)}(r) = \frac{1}{\sqrt{2\pi R\left(1-\dfrac{\sin 2x_n}{2x_n}\right)}} \frac{\sin(x_n r/R)}{r} .
\end{equation}
The lowest mode ($n=0$) has no nodes when $q>0$ and corresponds to the ground state, while modes with $n\geqslant 1$ have $n$ radial nodes.

To verify the spectral duality \eqref{eq:duality}, we study the convergence of the truncated series below
\begin{equation}
\label{eq:phiappRobin}
\phi_\mrm{app}(p,q|r) = \sum_{n=0}^{n_\mrm{max}} \frac{D\lambda_n^{(q)} \, a_n^{(q)} \, u_n^{(q)}(r)}{D\lambda_n^{(q)} + p} .
\end{equation}
Figure \ref{fig:duality} compares truncations at several values of $n_\mrm{max}$. 
The exact generalized Steklov result (Eq.~\eqref{eq:H_ball}) is shown as a thick black curve. 
Colored curves with symbols show truncations at the indicated $n_\mrm{max}$ values. 
The approximation~\eqref{eq:phiappRobin} improves overall as $n_\mrm{max}$ increases, although the convergence is nonuniform for $r \to 0$ and $r \to R$.

\section{Discussion and conclusions}
\label{sec:conclusion}

In this paper, we studied diffusion under joint stopping conditions, in which a diffusing particle in a bounded domain is terminated by either a bulk decay or a surface reaction, whichever occurs first. 
We showed that all statistics are governed by a single quantity $\phi(p,q|\x_0)$, which plays three equivalent roles: 
(i) the Laplace transform of the pure-Robin first-reaction time $\tau$ evaluated at the bulk decay rate $p$; 
(ii) the splitting probability $\p_{\x_0}(\tau < \delta)$ that surface reaction occurs before bulk decay; 
(iii) the effective surface weight $q\expect{L}$. 
In particular, the pathwise formulation reveals the identities: 
\begin{subequations}
\label{eq:balance}
\begin{align}
p \expect{T}(\x_0) &= \p_{\x_0}(\delta<\tau)
= \p_{\x_0}(\ell_\delta<\hat\ell),
\label{eq:balance_a} \\
q \expect{L}(\x_0) &= \p_{\x_0}(\tau<\delta)
= \p_{\x_0}(\hat\ell<\ell_\delta) ,
\label{eq:balance_b}
\end{align}
\end{subequations}
and the two mutually exclusive stopping channels have complementary probabilities. 
As a consequence, we obtain the exact balance relation $p\expect{T}+q\expect{L}=1$. 
Derived from the cumulative risk, the universal law $pT +qL \sim \mrm{Exp}(1)$ holds for an arbitrary bounded domain $\Omega$, starting position $\x_0$, and rates $p, q > 0$. 
Once $\phi(p,q|\x_0)$ is known, we obtain the joint generating function $W(\gamma,\eta|\x_0)$, and then we obtain all mixed moments. 
The Robin-Laplacian and generalized Steklov spectral viewpoints on diffusion-mediated reactions were discussed in the context of surface-hopping and conventional propagators \cite{Grebenkov20PRE, Grebenkov23PRE}. 
Under the present joint stopping condition, the two viewpoints converge on the splitting probability $\phi(p,q|\x_0)$, leading to the spectral duality~\eqref{eq:duality}. %of $\phi(p,q|\x_0)$ in terms of the Robin-Laplacian (Eq.~\eqref{eq:H_spectral}) and the generalized Steklov spectrum (Eq.~\eqref{eq:H_spectral2}). 
Explicit solutions for the three-dimensional ball allowed us to construct the heat map of $\phi(p,q|\x_0)$. %, identifying three regimes: lifetime-dominated, surface-dominated, and mixed. 
For starting points on the boundary, the constant contours $\phi_c$ cross over from $q \propto p$ to $q \propto p^{1/2}$. 
For uniformly distributed starting points, by contrast, a sufficiently fast bulk decay prevents a prescribed surface-reaction probability, even in the limit $q \to \infty$. 
For the mean stopping time $\expect{T}$, the heat map is separated into two different parts by the classical mean first-passage time $T_\mrm{D}$. 
Monte Carlo simulations based on the walk-on-spheres algorithm and the escape-from-a-layer method confirmed the theoretical predictions.

The framework also naturally extends in several directions. 
For experiments such as single-particle tracking \cite{saxton1997single, manzo2015review, simon2024guide},
if $p$ and $q$ can be calibrated independently, the sum rule~\eqref{eq:sum_rule} provides an indirect way to infer $\expect{L}$ from the mean stopping time via $\expect{L}=(1-p\expect{T})/q$, or from the splitting fraction via $\expect{L} = \phi/q$. 
This identity can generalize directly to multiple competing exponential stopping mechanisms acting on additive functionals of the trajectory. 
For spatially dependent bulk decay rates $p(\Omega_i)$ and surface reactivities $q(\Gamma_j)$, the total accumulated risk at the stopping time remains exponentially distributed with unit rate (see Appendix~\ref{app:cumulative_risk}). In contrast, the simple spectral shifts rely on spatially uniform $p$ and $q$, whereas spatially heterogeneous rates are no longer scalar shifts of the corresponding operators. 
An alternative extension consists in requiring both stopping conditions to be satisfied, i.e., $T^\prime = \max(\delta, \tau)$, and $L^\prime = \ell_{T^\prime}$. 
%Two thresholds $\delta$ and $\hat\ell$ remain independent, but both conditions must be satisfied for the process to end. 
Recent work extended the encounter-based framework to negative reactivities ($q < 0$) for the boundary branching process, where each encounter on the reactive surface creates new particles rather than absorbing them \cite{grebenkov2026geometric, grebenkov2026reaction, grebenkov2026population}. 
The quantity $\phi = q\expect{L}$ is not the splitting probability, and the sum rule or the exponential law break down. 
Similarly, we can consider the bulk proliferation with $p < 0$ and particle duplication at rate $|p|$. 
A zero growth rate of the mean population might be achieved when bulk decay exactly balances surface proliferation, or when the surface reaction balances the bulk proliferation.  
The interplay between growth and extinction ($p<0,q>0$ and $p>0,q<0$) opens a rich perspective.

\begin{acknowledgments}                                                               
The author thanks Prof.~Denis S.~Grebenkov for insightful discussions. 
The author used OpenAI's ChatGPT (GPT-5.6-sol) during the preparation of this manuscript to assist with language editing, manuscript organization, and exploratory checks of selected analytical derivations. All AI-assisted output was critically reviewed and independently verified by the author, who assumes full responsibility for the scientific content.
\end{acknowledgments}

\appendix

\section{Derivation of the mean identities}
\label{app:int}

Starting from Eq.~\eqref{eq:H_def}, we equivalently write 
\begin{align}                                                                      
\label{eq:H_alt}                                                                      
\phi(p,q|\x_0) &= \int_0^\infty \md t \, [f_T(t|\x_0) - f_{T,b}(t|\x_0)] \nonumber \\
&= 1 - p\int_0^\infty \md t \, e^{-pt} S_\tau(t|\x_0) ,
\end{align} 
which can be verified via integration by parts. 
The mean stopping time decomposes into the two contributions $I_1$ and $I_2$: 

\begin{align}
\expect{T} (\x_0) &= \int_0^\infty \md t \, t \, f_T(t|\x_0) = \int_0^\infty \md t \, e^{-pt} S_\tau(t|\x_0) \nonumber \\
&= \int_0^\infty \md t \, t \left[ e^{-pt} f_\tau(t|\x_0) + p\,e^{-pt} S_\tau(t|\x_0) \right] \nonumber \\
&= \underbrace{\int_0^\infty \md t \, t \, e^{-pt} f_\tau(t|\x_0)}_{I_1} + \underbrace{p \int_0^\infty \md t \, t \, e^{-pt} S_\tau(t|\x_0)}_{I_2} . 
\end{align}
Note that $I_1 = \int_0^\infty \md t \, t \, f_{T,s} (t|\x_0) = -\pdv{\phi}{p}$. 
On the other hand, we obtain
\begin{align}
I_2 &= \int_0^\infty \md t \, t \, f_{T,b} (t|\x_0) = - \int_0^\infty \md \left( e^{-pt} \right) \, t \, S_\tau(t|\x_0) \nonumber \\
&= \int_0^\infty \md \left[ t S_\tau(t|\x_0) \right] e^{-pt} - \left[ t \, e^{-pt} \, S_\tau(t|\x_0) \right]_0^\infty \nonumber \\
&= \int_0^\infty \left[ - t f_\tau(t|\x_0) \md t + S_\tau(t|\x_0) \md t \right] e^{-pt} \nonumber \\
&= \int_0^\infty \md t \, e^{-pt} S_\tau(t|\x_0) - I_1 . 
\end{align}
Therefore, we recover the equation \eqref{eq:Texpect}
\begin{align}
\expect{T}(\x_0) &= I_1 + I_2 = \int_0^\infty \md t \, e^{-pt} S_\tau(t|\x_0) \nonumber \\
&= \frac{1}{p} \left[ 1 - \int_0^\infty \md t \, e^{-pt} f_\tau(t|\x_0) \right] .
\end{align}

An analogous integration by parts for $L$ yields $\expect{L}$: 
\begin{align}
\expect{L} (\x_0) &= \int_0^\infty \md \ell \, \ell \, g_L(\ell|\x_0) = \int_0^\infty \md \ell \, e^{-q\ell} Q_\delta (\ell|\x_0) \nonumber \\
&= \int_0^\infty \md \ell \, \ell \left[ e^{-q\ell} \Psi(\ell,p|\x_0) + q\,e^{-q\ell} Q_\delta(\ell|\x_0) \right] \nonumber \\
&= \underbrace{\int_0^\infty \md \ell \, \ell \, e^{-q\ell} \Psi(\ell,p|\x_0)}_{I_3} + \underbrace{q \int_0^\infty \md \ell \, \ell \, e^{-q\ell} Q_\delta(\ell|\x_0)}_{I_4} . 
\end{align}
Similarly, 
\begin{align}
I_4 &= \int_0^\infty \md \ell \, \ell \, g_{L,s} (\ell|\x_0) = - \int_0^\infty \md \left( e^{-q\ell} \right) \, \ell \, Q_\delta(\ell|\x_0) \nonumber \\
&= \int_0^\infty \md \left[ \ell Q_\delta(\ell|\x_0) \right] e^{-q\ell} - \left[ \ell \, e^{-q\ell} \, Q_\delta(\ell|\x_0) \right]_0^\infty \nonumber \\
&= \int_0^\infty \left[ - \ell \Psi(\ell,p|\x_0) \md \ell + Q_\delta(\ell|\x_0) \md \ell \right] e^{-q\ell} \nonumber \\
&= \int_0^\infty \md \ell \, e^{-q\ell} Q_\delta(\ell|\x_0) - I_3 . 
\end{align}
Therefore, we recover the equation \eqref{eq:Lexpect}
\begin{align}
\expect{L}(\x_0) &= I_3 + I_4 = \int_0^\infty \md \ell \, e^{-q\ell} Q_\delta(\ell|\x_0) \nonumber \\
&= \frac{1}{q} \left[ 1 - \int_0^\infty \md \ell \, e^{-q\ell} \Psi(\ell,p|\x_0) \right] .
\end{align}

The quantities $I_1, I_2, I_3, I_4$ are unnormalized cause-specific moments. 
Conditional means $\expect{T}_\mrm{s}$, $\expect{T}_\mrm{b}$, $\expect{L}_\mrm{s}$, and $\expect{L}_\mrm{b}$ are shown in Eqs.~\eqref{eq:Tsmean}-\eqref{eq:Lbmean} in Appendix \ref{app:Wrecur}.

\section{Pure random lifetime ($q=0$) and Steklov expansion}
\label{app:pure_lifetime}

The boundary local time $\ell_\delta$ obtained at a random lifetime $\delta$ obeys the well-known distribution \cite{Ye25JCP}, such that 
$\Psi(\ell,p|\x_0) = p \tilde{\rho}(\ell,p|\x_0)$. This describes reflected Brownian motion with a Neumann boundary condition. 
The generalized probability density $\rho(\ell,t|\x_0)$ of the boundary local time $\ell_t$ reads in the Laplace domain as \cite{Grebenkov2019, Ye25JCP}
\begin{align}
\tilde{\rho}(\ell,p|\x_0) &= \int_0^\infty \md t \, e^{-p t} \rho(\ell,t|\x_0) \nonumber \\
&= \tilde{S}_\infty (p|\x_0) \delta_0(\ell) + \sum_{k=0}^\infty e^{-\mu_k^{(p)} \ell} c_k^{(p)}(\x_0) , 
\end{align}
where $\delta_0(z)$ is the Dirac delta distribution, 
\begin{equation}
c_k^{(p)}(\x_0) = \frac{[V_k^{(p)}(\x_0)]^\ast}{D} \int_\Omega \md \x \, V_k^{(p)}(\x) , 
\end{equation}
and $\tilde{S}_\infty (p|\x_0)$ is the Laplace-transformed survival probability in the presence of a perfect absorbing boundary (Dirichlet boundary condition, i.e., $q=\infty$). 
We can consider the statistics of $\ell_\delta$ directly in the Laplace domain without Laplace transform inversion by introducing
\begin{equation}
\Psi(\ell,p|\x_0) = \int_0^\infty \md t \, p \, e^{-pt} \, \rho(\ell,t|\x_0) = p \, \tilde{\rho}(\ell,p|\x_0) . 
\end{equation}

In Eq.~\eqref{eq:Psibk}, we introduce $b_k^{(p)}$ as the surface integral on $\Gamma$ of $V_k^{(p)}(\x)$. 
By the divergence theorem, we obtain
\begin{align} 
&p \int_\Omega \md x \, V_k^{(p)} = D \int_\Omega \md x \,\Delta V_k^{(p)} \nonumber \\
&= D \int_\Gamma \md S \, \partial_n V_k^{(p)} = D \mu_k^{(p)} \, b_k^{(p)} ,
\end{align}
thus 
\begin{equation}
p \, c_k^{(p)}(\x_0) = [V_k^{(p)}(\x_0)]^\ast \, \mu_k^{(p)} \, b_k^{(p)} . 
\end{equation}

\section{Multiple clocks and cumulative risk}
\label{app:cumulative_risk}

The competition among random lifetimes has been investigated in \cite{Meerson15}, while the competition among boundary local times was investigated in \cite{Grebenkov20JSM}. 
We extend the identity to multiple competing clocks including both the ordinary diffusion time and the boundary local time.

Consider a domain $\Omega$ separated into $N$ disjoint subdomains $\Omega_1,\dots,\Omega_N$ (i.e., $\Omega = \bigcup_{i=1}^N \Omega_i$), and a boundary $\Gamma$ divided into $M$ disjoint subsurfaces $\Gamma_1,\dots,\Gamma_M$ (i.e., $\Gamma = \bigcup_{j=1}^M \Gamma_j$). 
In each $\Omega_i$, the particle undergoes bulk decay with rate $p_i>0$, such that the independent thresholds $\delta_i \sim \mrm{Exp}(p_i)$. 
On each $\Gamma_j$, the particle reacts with rate $q_j>0$, such that  
the independent thresholds $\hat\ell_j \sim \mrm{Exp}(q_j)$. 
Define the occupation time of $\Omega_i$ and the boundary local time accumulated on $\Gamma_j$ up to time $t$ as $\mathcal T_i(t)$ and $\ell_j(t)$, respectively:
\begin{subequations}
\begin{align}
\mathcal{T}_i(t) &=\int_0^t \md s \, \mathbf 1_{\Omega_i}(\X_s)\, , 
\\
\ell_j (t) &= \lim_{\varepsilon \to 0} \frac{D}{\varepsilon} \int_0^t \md s \, \mathbf{1}_{\Gamma_j^{(\varepsilon)}}(\X_s) ,
\end{align}
\end{subequations}
where $\Gamma_j^{(\varepsilon)}$ is the boundary layer of thickness $\varepsilon$. 
The stopping times on each part are
\begin{subequations}
\begin{align}
\sigma_i &= \inf\{t>0:\mathcal T_i(t)>\delta_i\},
\\
\tau_j &= \inf\{t>0:\ell_j(t)>\hat\ell_j\},
\end{align}
\end{subequations}
and the global stopping time is
\begin{equation}
\mathcal T=\min\{\sigma_1,\dots,\sigma_N,\tau_1,\dots,\tau_M\}.
\label{eq:global_stop}
\end{equation}
We denote the stopped occupation and local times by
$T_i=\mathcal T_i(\mathcal T)$ and $L_j=\ell_j(\mathcal T)$.

Then we define the \emph{cumulative risk} process
\begin{equation}
\mathcal{R}_t=\sum_{i=1}^N p_i \mathcal T_i(t)+\sum_{j=1}^M q_j \ell_j(t).
\label{eq:cumrisk_app}
\end{equation}
Conditioned on the full trajectory $\X_t$, the thresholds are independent, so
for any $t>0$,
\begin{align}
\p(\mathcal T>t\,|\,\X_t)
&=\prod_{i=1}^N\p\bigl(\mathcal T_i(t)<\delta_i\bigr)
  \prod_{j=1}^M\p\bigl(\ell_j(t)<\hat\ell_j\bigr) \nonumber\\
&=\exp\!\left[-\sum_{i=1}^N p_i \mathcal T_i(t)
                -\sum_{j=1}^M q_j \ell_j(t)\right]
 =e^{-\mathcal{R}_t}.
\end{align}
The process $\mathcal{R}_t$ is continuous and strictly increasing for every trajectory, with $\mathcal{R}_0=0$ and $\mathcal{R}_t\to\infty$ as $t\to\infty$. Hence for each $z>0$ there exists
a unique $t_z$ such that $\mathcal{R}_{t_z}=z$.  Consequently,
\begin{equation}
\p(\mathcal{R}_{\mathcal T}>z | \X_t ) =\p(\mathcal T>t_z | \X_t ) =e^{-\mathcal{R}_{t_z}}=e^{-z}.
\end{equation}
The term $e^{-z}$ is independent of the trajectory, so averaging over the reflected path yields $\p(\mathcal{R}_{\mathcal T}>z)=e^{-z}$, i.e.,
\begin{equation}
\mathcal{R}_{\mathcal T} =\sum_{i=1}^N p_i T_i+\sum_{j=1}^M q_j L_j \sim \mrm{Exp}(1) .
\label{eq:universal_app}
\end{equation}
Since $\mathcal{R}_{\mathcal T}$ has unit mean, we immediately obtain the generalized
first-moment identity
\begin{equation}
\sum_{i=1}^N p_i\langle T_i\rangle
+\sum_{j=1}^M q_j\langle L_j\rangle=1.
\label{eq:generalized_first_moment}
\end{equation}
More precisely, we define $\phi_j=\p(\mathcal{T} = \tau_j)=q_j \expect{L_j}$ and $\varphi_i=\p(\mathcal{T} = \sigma_i)=p_i \expect{T_i}$. These probabilities sum to one. 
When $N=M=1$, Eq.~\eqref{eq:generalized_first_moment} reduces to Eq.~\eqref{eq:sum_rule} as expected.
Moreover, all higher moments of the total risk are those of $\mrm{Exp}(1)$.

For more general thresholds 
$\p\{ \delta > t \} = e^{-\mathscr{P}(t)}$ and $\p \{ \hat\ell > \ell \}  = e^{-\mathscr{Q}(\ell)}$, where $\mathscr{P}(t)$ and $\mathscr{Q}(\ell)$ are arbitrary continuous and nondecreasing functions, satisfying $\mathscr{P}(0) = \mathscr{Q}(0) = 0$ and $\lim_{t\to\infty} \mathscr{P}(t) = \lim_{\ell\to\infty} \mathscr{Q}(\ell) = \infty$, 
we can follow the same procedure to prove the universal exponential law:  
\begin{equation}
\mathscr{P}(T) + \mathscr{Q}(L) \sim \mrm{Exp}(1) . 
\end{equation}
Moreover, we can prove $\E[\mathscr{P}(T)] = \p(\delta < \tau) = 1 - \phi$ and $\E[\mathscr{Q}(L)] = \p(\tau < \delta) = \phi$. Therefore,  
\begin{equation}
\E[\mathscr{P}(T)] + \E[\mathscr{Q}(L)] = 1 .
\end{equation}
Again, multiple different functions $\mathscr{P}_i(t)$ and $\mathscr{Q}_j(\ell)$ lead to
\begin{subequations}
\begin{align}
\sum_{i=1}^N \mathscr{P}_i(T_i) + \sum_{j=1}^M \mathscr{Q}_j(L_j) \sim \mrm{Exp}(1) , \\
\sum_{i=1}^N \E [\mathscr{P}_i(T_i)] + \sum_{j=1}^M \E [\mathscr{Q}_j(L_j)] = 1.
\end{align}
\end{subequations}

\section{Conditional statistics}
\label{app:Wrecur}

Cause-specific transforms satisfy
\begin{subequations}
\begin{align}
(D \Delta-p-\gamma)W_\mrm{b}(\gamma,\eta|\x_0) &= -p \quad (\x_0\in\Omega) , \\
(\partial_n+q+\eta)W_\mrm{b}(\gamma,\eta|\x_0) &= 0 \qquad (\x_0\in\Gamma) , \\
(D \Delta-p-\gamma)W_\mrm{s}(\gamma,\eta|\x_0) &= 0 \qquad (\x_0\in\Omega) , \\
(\partial_n+q+\eta)W_\mrm{s}(\gamma,\eta|\x_0) &= q \qquad (\x_0\in\Gamma) . 
\end{align}
\end{subequations}
When $\gamma \to 0$ and $\eta \to 0$, we recover the relation $W_\mrm{s}(0,0|\x_0) = \E_{\x_0} \left[\mathbf 1_{\tau<\delta}\right] = \p_{\x_0}(\tau<\delta) = \phi(p,q|\x_0)$. As a result, we obtain
\begin{subequations}
\begin{align}
(D \Delta-p) \phi(p,q|\x_0) &= 0 \qquad (\x_0\in\Omega) , \\
(\partial_n+q) \phi(p,q|\x_0) &= q \qquad (\x_0\in\Gamma) . 
\end{align}
\end{subequations}
The joint Laplace transform $W(\gamma,\eta|\x_0)$ satisfies: \begin{subequations}
\label{eq:moment_formulas}
\begin{align}
(D \Delta-p-\gamma)W(\gamma,\eta|\x_0) &= -p \quad (\x_0\in\Omega) , \\
(\partial_n+q+\eta)W(\gamma,\eta|\x_0) &= q \qquad (\x_0\in\Gamma) , 
\end{align}
\end{subequations}
Define $W_{m,n}(\x_0) = \expect{T^m L^n} (\x_0)$, we can also produce the full moment hierarchy of $(T,L)$:
\begin{subequations}
\begin{align}
\label{eq:recurb}
(D \Delta - p) W_{m,n}(\x_0) &= - m W_{m-1,n}(\x_0) \quad (\x_0\in\Omega), \\
\label{eq:recurs}
(\partial_n + q) W_{m,n}(\x_0) &= n W_{m,n-1}(\x_0) \qquad (\x_0\in\Gamma) , 
\end{align}
\end{subequations}
where $m,n \geqslant 0$, $m+n \geqslant 1$, and $W_{0,0}=1$. 
Terms with negative indices are zero. 
Increasing the time order $m$ generates an interior source (Eq.~\eqref{eq:recurb}), whereas increasing the boundary local time order $n$ generates a boundary source (Eq.~\eqref{eq:recurs}). 
\begin{subequations}
\begin{align}
\langle T^m\rangle (\x_0)&= (-1)^{m-1}\,m\,\partial_p^{m-1} \expect{T} (\x_0),
\label{eq:T_moment} \\
\langle L^n\rangle (\x_0)&= (-1)^{n-1}\,n\,\partial_q^{n-1} \expect{L} (\x_0),
\label{eq:L_moment} \\
\langle T^m L^n\rangle (\x_0)&= (-1)^{m+n-1} \left(m\,\partial_p^{m-1}\partial_q^n \expect{T} (\x_0)\right. \nonumber \\ & \qquad \qquad \left. + n\,\partial_p^m\partial_q^{n-1} \expect{L} (\x_0)\right) .
\end{align}
\end{subequations}
Recall that 
\begin{equation}
\label{eq:m_first}
\expect{T}(\x_0) = \frac{1-\phi(p,q|\x_0)}{p}, \quad
\expect{L}(\x_0) = \frac{\phi(p,q|\x_0)}{q} . 
\end{equation}

Apart from these overall mean values in the first order, 
cause-specific transforms also yield conditional means for trajectories stopped by bulk decay or surface reaction:
\begin{subequations}
\begin{align}
\expect{T}_\mrm{s}(\x_0) &= \E_{\x_0}[T|\tau<\delta] = - \partial_\gamma \left. \frac{W_\mrm{s}(\gamma,\eta|\x_0)}{\phi(p,q|\x_0)} \right\vert_{\gamma=\eta=0} \nonumber \\ &= - \frac{\phi_p(p,q|\x_0)}{\phi(p,q|\x_0)} , \label{eq:Tsmean} \\
\expect{T}_\mrm{b}(\x_0) &= \E_{\x_0}[T|\delta<\tau] = - \partial_\gamma \left. \frac{W_\mrm{b}(\gamma,\eta|\x_0)}{1-\phi(p,q|\x_0)} \right\vert_{\gamma=\eta=0} \nonumber \\ &= \frac{1}{p} + \frac{\phi_p(p,q|\x_0)}{1-\phi(p,q|\x_0)} , \label{eq:Tbmean} \\
\expect{L}_\mrm{s}(\x_0) &= \E_{\x_0}[L|\tau<\delta] = - \partial_\eta \left. \frac{W_\mrm{s}(\gamma,\eta|\x_0)}{\phi(p,q|\x_0)} \right\vert_{\gamma=\eta=0} \nonumber \\ &= \frac{1}{q} - \frac{\phi_q(p,q|\x_0)}{\phi(p,q|\x_0)} , \label{eq:Lsmean} \\
\expect{L}_\mrm{b}(\x_0) &= \E_{\x_0}[L|\delta<\tau] = - \partial_\eta \left. \frac{W_\mrm{b}(\gamma,\eta|\x_0)}{1-\phi(p,q|\x_0)} \right\vert_{\gamma=\eta=0} \nonumber \\ &= \frac{\phi_q(p,q|\x_0)}{1-\phi(p,q|\x_0)} \label{eq:Lbmean} , 
\end{align}
\end{subequations}
where $\phi_p\equiv\partial_p\phi$ and $\phi_q\equiv\partial_q\phi$ are  derivatives (Eq.~\eqref{eq:phi_derivatives}).

The second moments read: 
\begin{subequations}
\begin{align}
\label{eq:m_secondT}
\langle T^2\rangle(\x_0) &= \frac{2[1-\phi(p,q|\x_0)]}{p^2}+\frac{2\phi_p(p,q|\x_0)}{p}, \\
\label{eq:m_secondL}
\langle L^2\rangle(\x_0) &= \frac{2\phi(p,q|\x_0)}{q^2}-\frac{2\phi_q(p,q|\x_0)}{q}, \\
\label{eq:m_cross}
\langle TL\rangle(\x_0) &= \frac{\phi_q(p,q|\x_0)}{p}-\frac{\phi_p(p,q|\x_0)}{q} .
\end{align}
\end{subequations}
Generated by the same trajectory,  $T$ and $L$ are generally correlated \cite{Ye26JCP}. 
We obtain the correlation coefficient $C(p,q|\x_0)$: 
\begin{align}
C(p,q|\x_0) &= \frac{\langle TL\rangle-\expect{T}\expect{L}}{\sqrt{\Var(T)\,\Var(L)}} (p,q|\x_0) \nonumber \\
&= \frac{-p \phi _p+q \phi _q+(\phi -1) \phi }{\sqrt{\left( 1 - \phi^2 + 2 p \phi_p\right) \left(2\phi - \phi^2 - 2 q \phi_q \right)}} (p,q|\x_0), 
\label{eq:Cqdef}
\end{align}
which requires nonzero variances. In singular limits $p\to~\infty$ or $q \to \infty$, one needs a limiting interpretation.

\section{Explicit solutions for additional geometries}
\label{app:moredomains}

\subsection{One-dimensional interval}

We consider the one-dimensional interval $\Omega = [0,R] \subset \mathbb{R}$, with a Robin boundary condition at  $x=0$ and a Neumann boundary condition at $x=R$. 
Because the reactive boundary consists of the single endpoint, only one Steklov eigenmode contributes to $\phi(p,q|x)$: 
\begin{subequations}
\begin{align}
V_{0;\mrm{1d}}^{(p)}(x) &= \frac{\cosh[\alpha(R-x)]}{\cosh(\alpha R)} , \\
\mu_{0;\mrm{1d}}^{(p)} &= \alpha \tanh(\alpha R) , 
\end{align}
\end{subequations}
where $\alpha = \sqrt{p/D}$. 
We obtain the splitting probability
\begin{equation}
\phi_\mrm{1d}(p,q|x) = \frac{q}{\alpha \tanh(\alpha R) + q} \frac{\cosh[\alpha(R-x)]}{\cosh(\alpha R)} . 
\end{equation}
When $R\to\infty$, we obtain
\begin{equation}
\phi_{\mrm{1d},\infty}(p,q|x) = \frac{q}{\alpha+q} e^{- \alpha x} . 
\end{equation}
Because one-dimensional Brownian motion is recurrent, the probability to return to the point $x=0$ equals 1. 
In other words, in the limit $p\to0$, $\phi_\mrm{1d}(p,q|x) \to 1$ and $\phi_{\mrm{1d},\infty}(p,q|x) \to 1$. 
However, for the latter case, the mean stopping time diverges as $p\to0$:
\begin{equation}
\expect{T}_{\mrm{1d},\infty} (p,q|x) \approx \frac{x+1/q}{\sqrt{pD}} . 
\end{equation}

\subsection{Two-dimensional disk}
We consider the two-dimensional disk $\Omega = \{ |\x| < R \} \subset \mathbb{R}^2$, with a Robin boundary condition at $r=R$. 
Although the complete Steklov spectrum contains all angular modes, only the rotationally invariant mode contributes to $\phi(p,q|r)$: 
\begin{subequations}
\begin{align}
V_{0;\mrm{2d}}^{(p)}(r) &= \frac{1}{\sqrt{2\pi R}} \frac{I_0(\alpha r)}{I_0(\alpha R)} , \\
\mu_{0;\mrm{2d}}^{(p)} &= \alpha \frac{I_1(\alpha R)}{I_0(\alpha R)} , 
\end{align}
\end{subequations}
where $\alpha = \sqrt{p/D}$ and $I_n(z)$ denotes the modified Bessel function of the first kind of order $n$. 
We obtain the splitting probability:
\begin{equation}
\phi_\mrm{2d}(p,q|r) = \frac{q I_0(\alpha r)}{\alpha I_1(\alpha R) + q I_0(\alpha R)}. 
\end{equation}
In this bounded domain, the mean stopping time is finite.

\subsection{Two-dimensional circular annulus}
We consider the two-dimensional concentric circular annulus $\Omega = \{ R_1 < |\x| < R_2 \} \subset \mathbb{R}^2$, with a Robin boundary condition  at $r = R_1$ and a Neumann boundary condition at $r = R_2$. 
Note that the outward normal at the inner reactive boundary $r = R_1$ points toward decreasing $r$. 
Only the rotationally invariant mode contributes to $\phi(p,q|r)$: 
\begin{subequations}
\begin{align}
V_{0;\mrm{2d,ext}}^{(p)}(r)&=\frac{1}{\sqrt{2\pi R_1}} \frac{I_1\left(\alpha R_2\right) K_0\left(\alpha r\right)+K_1\left(\alpha R_2\right) I_0\left(\alpha r\right)}{I_1\left(\alpha R_2\right) K_0\left(\alpha R_1\right)+K_1\left(\alpha R_2\right) I_0\left(\alpha R_1\right)} , \\
\mu_{0;\mrm{2d,ext}}^{(p)} &= \frac{\alpha \left(I_1\left(\alpha R_2\right) K_1\left(\alpha R_1\right)-K_1\left(\alpha R_2\right) I_1\left(\alpha R_1\right)\right)}{I_1\left(\alpha R_2\right) K_0\left(\alpha R_1\right)+K_1\left(\alpha R_2\right) I_0\left(\alpha R_1\right)}, 
\end{align}
\end{subequations}
where $\alpha = \sqrt{p/D}$ and $K_n(z)$ is the modified Bessel function of the second kind. 
We obtain the splitting probability
\begin{equation}
\phi_\mrm{2d,ext}(p,q|r) = \frac{q \, b_{0;\mrm{2d,ext}}^{(p)}}{\mu_{0;\mrm{2d,ext}}^{(p)} + q} \, V_{0;\mrm{2d,ext}}^{(p)}(r) , 
\end{equation}
where $b_{0;\mrm{2d,ext}}^{(p)} = \sqrt{2\pi R_1}$. 
In the limit $R_2\to\infty$, $I_1\left(\alpha R_2\right) \to \infty$ and $K_1\left(\alpha R_2\right) \to 0$. As a consequence, we obtain 
\begin{subequations}
\begin{align}
V_{0;\mrm{2d,ext},\infty}^{(p)}(r) &= \frac{1}{\sqrt{2\pi R_1}} \frac{K_0\left(\alpha r\right)}{K_0\left(\alpha R_1\right)} , \\
\mu_{0;\mrm{2d,ext},\infty}^{(p)} &= \frac{\alpha K_1\left(\alpha R_1\right)}{K_0\left(\alpha R_1\right)} . 
\end{align}
\end{subequations}
Therefore, 
\begin{equation}
\phi_{\mrm{2d,ext},\infty}(p,q|r) = \frac{q K_0\left(\alpha r\right)}{q K_0\left(\alpha R_1\right)+\alpha K_1\left(\alpha R_1\right)} . 
\end{equation}
Since planar Brownian motion is also recurrent, the exterior Brownian particle starting at arbitrary position hits the inner circle $r=R_1$ almost surely. 
In other words, in the limit $p\to0$, $\phi_\mrm{2d,ext}(p,q|r) \to 1$ and $\phi_{\mrm{2d,ext},\infty}(p,q|r) \to 1$. 
However, for the latter case, the mean stopping time diverges as $p\to0$: 
\begin{equation}
\expect{T}_{\mrm{2d,ext},\infty} (p,q|r) \approx 
\frac{\log(r/R_1)+1/(qR_1)}{p[\log(2/(\alpha R_1))- \gamma_\mrm{E} +1/(qR_1)]} ,
\end{equation}
where $\gamma_\mrm{E}$ refers to the Euler-Mascheroni constant. 
This expression can be obtained via the asymptotic behaviors $K_0(z) \approx \log (2/z)- \gamma_\mrm{E}$ and $K_1(z) \approx 1/z$ when $z\to0$.

\subsection{Three-dimensional spherical shell}

We consider the three-dimensional concentric spherical shell $\Omega = \{ R_1 < |\x| < R_2 \} \subset \mathbb{R}^3$, with a Robin boundary condition  at $r = R_1$ and a Neumann boundary condition at $r = R_2$. Note that the outward normal at the inner reactive boundary $r = R_1$ points toward decreasing $r$. 
Only the rotationally invariant Steklov eigenmode contributes to $\phi(p,q|r)$: 
\begin{subequations}
\begin{align}
V_{0;\mrm{3d,ext}}^{(p)}(r) &= \frac{1}{\sqrt{4\pi R_1^2}} \frac{i_1\left(\alpha R_2\right) k_0\left(\alpha r\right)+k_1\left(\alpha R_2\right) i_0\left(\alpha r\right)}{i_1\left(\alpha R_2\right) k_0\left(\alpha R_1\right)+k_1\left(\alpha R_2\right) i_0\left(\alpha R_1\right)} , \\
\mu_{0;\mrm{3d,ext}}^{(p)} &= \frac{\alpha \left(i_1\left(\alpha R_2\right) k_1\left(\alpha R_1\right)-k_1\left(\alpha R_2\right) i_1\left(\alpha R_1\right)\right)}{i_1\left(\alpha R_2\right) k_0\left(\alpha R_1\right)+k_1\left(\alpha R_2\right) i_0\left(\alpha R_1\right)}, 
\end{align}
\end{subequations}
where $\alpha = \sqrt{p/D}$, $i_n(z)$ and $k_n(z)$ are the modified spherical Bessel functions of the first and the second kind, respectively. 
To be exact, $i_0(z) = \sinh(z)/z$, $i_1(z) = \cosh(z)/z-\sinh(z)/z^2$, $k_0(z) = e^{-z}/z$, and $k_1(z) = (1+1/z) \, e^{-z}/z$.

We obtain the splitting probability
\begin{equation}
\phi_\mrm{3d,ext}(p,q|r) = \frac{q \, b_{0;\mrm{3d,ext}}^{(p)}}{\mu_{0;\mrm{3d,ext}}^{(p)} + q} V_{0;\mrm{3d,ext}}^{(p)}(r) ,  
\end{equation}
where $b_{0;\mrm{3d,ext}}^{(p)} = \sqrt{4\pi R_1^2}$. 
In the limit $R_2\to\infty$, we obtain
\begin{equation}
\phi_{\mrm{3d,ext},\infty}(p,q|r) = \frac{q R_1^2 e^{\alpha (R_1-r)}}{r (\alpha R_1 + qR_1 +1)} , 
\end{equation}
which is the exterior-sphere counterpart of Eq.~\eqref{eq:phipapp}. The curvature term changes from $-1$ to $+1$ because the reactive boundary is viewed from the exterior. 
Since three-dimensional Brownian motion is transient, the exterior Brownian particle starting at $r>R_1$ hits the sphere with probability $R_1/r$. With finite surface reactivity $q$, the eventual reaction probability in the limit $p=0$ is $\phi(0,q|r) = (R_1/r) [qR_1/(1+qR_1)]$, which is smaller because the particle may escape after reflected encounters. Denote $\chi$ the probability for the escape event without surface reaction, and we have $q \expect{L} + \chi = 1$. Therefore, we emphasize the bounded domain or the nonzero decay rate $p$ to ensure the sum rule~\eqref{eq:sum_rule}.

\section{Moments and correlations in the ball}
\label{app:H_derivatives_ball}               
                                                                               
%% Figure <T>(p,q|R) & <L>(p,q|R)
\begin{figure}[t!]
    \includegraphics[width=\linewidth]{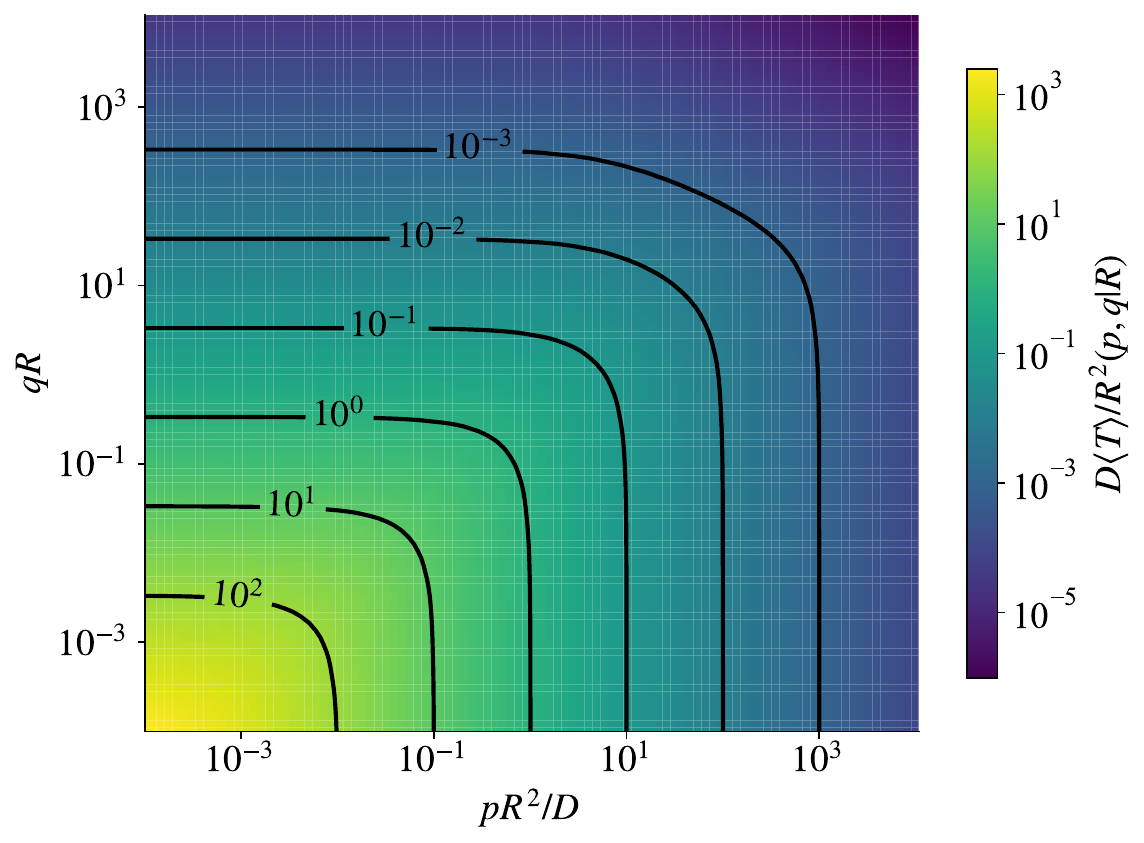}
    \rlap{\hspace{-0.5\linewidth}\raisebox{-0.0\linewidth}{\textbf{(a)}}}
\\
    \includegraphics[width=\linewidth]{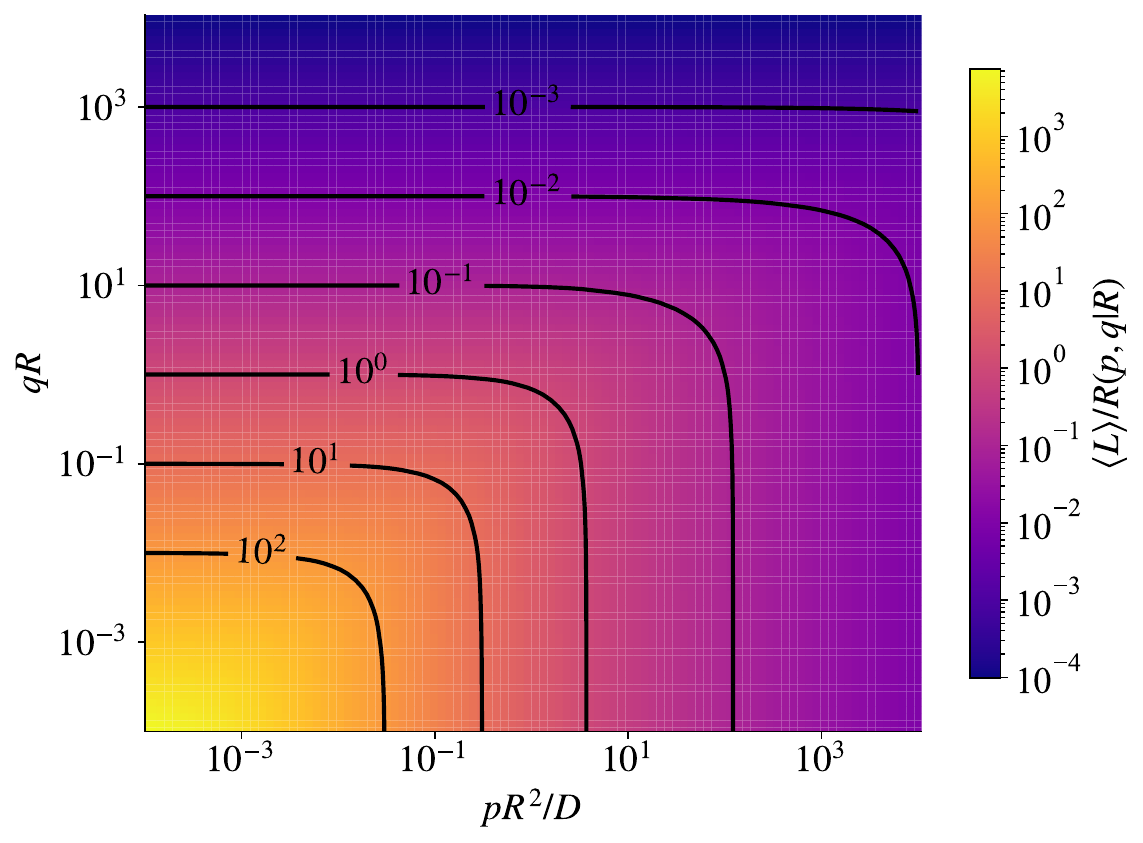}
    \rlap{\hspace{-0.5\linewidth}\raisebox{-0.0\linewidth}{\textbf{(b)}}}
\caption{Heat maps for uniformly distributed starting points on the reactive boundary of a three-dimensional ball ($R=1$ and $D=1$) over the $(pR^2/D,qR)$ plane in log-log scale. 
\textbf{(a)} Mean stopping time $\expect{T}(p,q|R)$. 
Black curves indicate contour lines for $\expect{T}(p,q|R) = 10^{-3}$, $10^{-2}$, $10^{-1}$, $10^0$, $10^{1}$, $10^{2}$. 
\textbf{(b)} Mean acquired boundary local time $\expect{L}(p,q|R)$. 
Black curves indicate contour lines for $\expect{L}(p,q|R) = 10^{-3}$, $10^{-2}$, $10^{-1}$, $10^0$, $10^{1}$, $10^{2}$. 
}
\label{fig:meanR}
\end{figure}

With $\phi(p,q|R)$ shown in Eq.~\eqref{eq:phiR}, we can derive the first-order averages for the starting point on the reactive surface. 
The heat maps of $\expect{T}(p,q|R)$ and $\expect{L}(p,q|R)$ are shown in Fig.~\ref{fig:meanR}(a) and Fig.~\ref{fig:meanR}(b), respectively. 
Unlike Fig.~\ref{fig:meano}(a), contour lines exhibit no contour-topology change in Fig.~\ref{fig:meanR}(a), whereas Fig.~\ref{fig:meanR}(b) is similar to Fig.~\ref{fig:meano}(b). 
Indeed, Eq.~\eqref{eq:MFRT} implies the mean first-reaction time as $(R^2 - r^2)/(6D) + R/(3Dq)$. 
Only at $r=R$ the first term $(R^2 - r^2)/(6D)$ vanishes, thus we can expect the contour-topology change for any $r<R$.

For uniform starting points in the ball, $\phi(p,q|\circ)$ is given by Eq.~\eqref{eq:phiunif}. 
Let $b=\alpha R$, the second moment of the stopping time shown in Eq.~\eqref{eq:m_secondT} can be written as
\begin{equation}
\expect{T^2}(p,q|\circ)
= \frac{R^4}{D^2}
\frac{\zeta_2+\zeta_3 qR +\zeta_4 (qR)^2}
{b^6(\zeta_1+ qR \tanh b)^2},
\label{eq:meanT2}
\end{equation}
where
\begin{subequations}
\begin{align}
\zeta_1 &= b-\tanh b , \\
\zeta_2 &=2b^2\zeta_1^2, \\
\zeta_3 &=4\zeta_1\left(b^2\tanh b-3\zeta_1\right), \\
\zeta_4 &=(5b^2+12)(\tanh b)^2-9b\tanh b-3b^2.
\end{align}
\end{subequations}
When $q\to0$, Eq.~\eqref{eq:meanT2} reduces to $2/p^2$, as required for $\delta \sim \mrm{Exp}(p)$. 
Similarly, the second moment of the stopped boundary local time shown in Eq.~\eqref{eq:m_secondL} reads
\begin{equation}
\expect{L^2}(p,q|\circ) = \frac{6R^2\zeta_1\tanh b}{b^2(\zeta_1+qR \tanh b)^2}.
\label{eq:meanL2}
\end{equation}
In the limit $p\to0$, Eq.~\eqref{eq:meanL2} reduces to $2/q^2$, as required for $\hat\ell \sim \mrm{Exp}(q)$. 

%% C(p,q|\circ)
\begin{figure}[t!]                                                                  
\includegraphics[width=\linewidth]{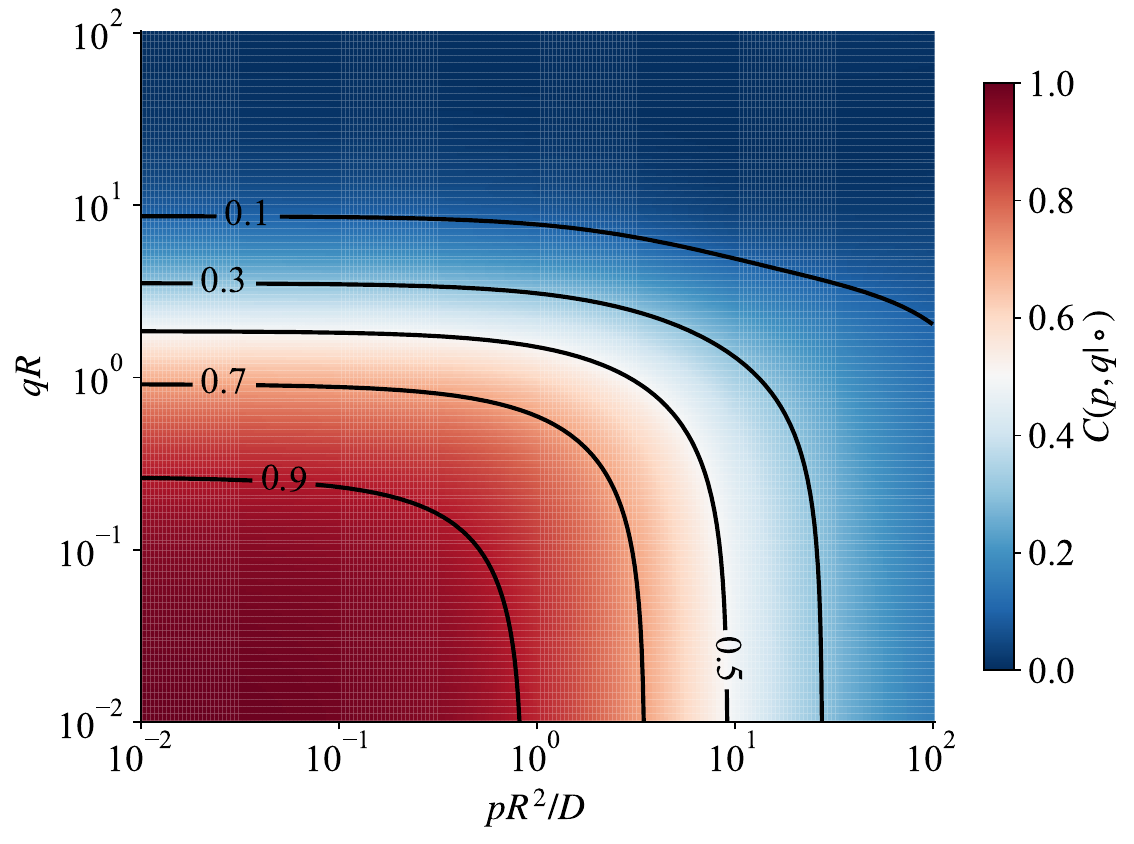} 
\caption{Heat map of the correlation coefficient $C(p,q|\circ)$ between $T$ and $L$ for uniformly distributed starting points in the ball ($R=1$, $D=1$) over the $(pR^2/D,qR)$ plane in log-log scale. 
Black curves exhibit contour lines at $C(p,q|\circ) = 0.1$, $0.3$, $0.5$, $0.7$, $0.9$. 
\label{fig:corr}}
\end{figure}

For a particle whose starting point is uniformly distributed in the unit ball ($R=1$, $D=1$), the correlation coefficient $C(p,q|\circ)$ is given by
\begin{equation}
C(p,q|\circ) = \sqrt{\frac{3}{8}} \frac{\xi_1(p) + \xi_2(p) q}{\sqrt{\xi_3(p) \xi_4(p) (\xi_5(p) + \xi_6(p) q + \xi_7(p) q^2)}} ,
\label{eq:corr}
\end{equation}
where
\begin{subequations}
\begin{align}
\xi_1(p) &= 4 p \xi_3^2 , \\ 
\xi_2(p) &= 2 p (p+3)-\sqrt{p} (p+12) \sinh(2 \sqrt{p}) \nonumber \\ & \quad +6 (p+1) \cosh (2 \sqrt{p})-6 , \\
\xi_3(p) &= \sqrt{p} \cosh \left(\sqrt{p}\right)-\sinh \left(\sqrt{p}\right) , \\ 
\xi_4(p) &= (2 p+3) \sinh \left(\sqrt{p}\right)-3 \sqrt{p} \cosh \left(\sqrt{p}\right) , \\
\xi_5(p) &= 2 p^2 \xi_3^2 , \\ 
\xi_6(p) &= 2 p \left(-2 p+(p+6) \sqrt{p} \sinh \left(2 \sqrt{p}\right) \right. \nonumber \\ & \quad \left. -(4 p+3) \cosh \left(2 \sqrt{p}\right)+3\right) , \\
\xi_7(p) &= -p (7 p+15)-3 (p-6) \sqrt{p} \sinh \left(2 \sqrt{p}\right)\nonumber \\ & \quad +((p-3) p-9) \cosh \left(2 \sqrt{p}\right)+9 . 
\end{align}
\end{subequations}

For uniformly distributed starting points in the ball, the correlation remains positive over the explored parameter range. It approaches unity only in the limit of small $p$ and small $q$, whereas increasing $p$ and $q$ generally weaken the correlation.                                                                                                                             
Strong correlations emerge in the slow surface-reaction and well-mixed regime ($p\to0, q\to0$), where the accumulated boundary local time is asymptotically proportional to the diffusion time.

The splitting probability $\phi(p,q|\x_0)$ and the correlation coefficient $C(p,q|\x_0)$ characterize different aspects of the competition. Comparable probabilities of the two stopping mechanisms, e.g., $\phi(p,q|\x_0) \approx 1/2$, do not imply weak correlations. When both stopping mechanisms are slow compared with diffusive mixing, the accumulated boundary local time $\ell_t$ is approximately proportional to the time $t$, $\ell_t \approx D|\Gamma|t/|\Omega|$, leading to strong correlations between $T$ and $L$.

\section{Additional Monte Carlo validation}
\label{app:otherphir}

\begin{figure}[t!]
\includegraphics[width=\linewidth]{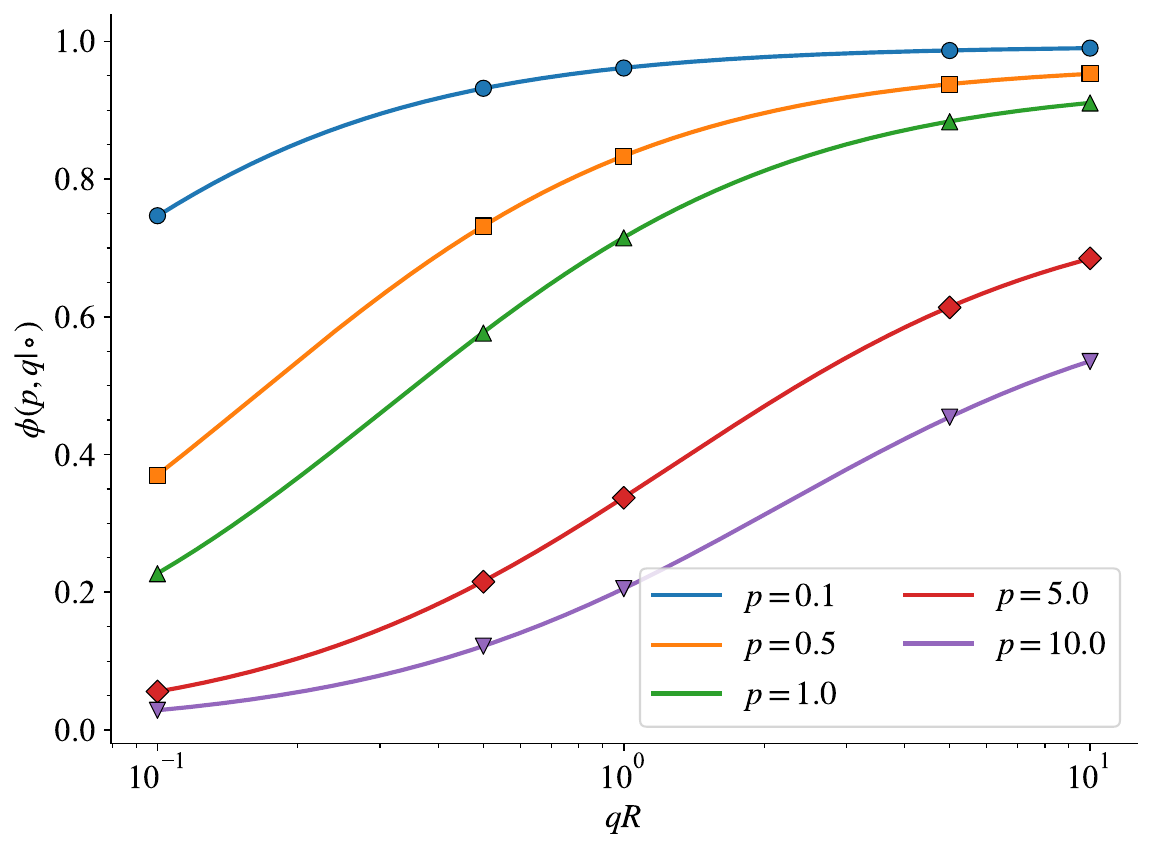}
\rlap{\hspace{-0.5\linewidth}\raisebox{-0.0\linewidth}{\textbf{(a)}}}
\\
\includegraphics[width=\linewidth]{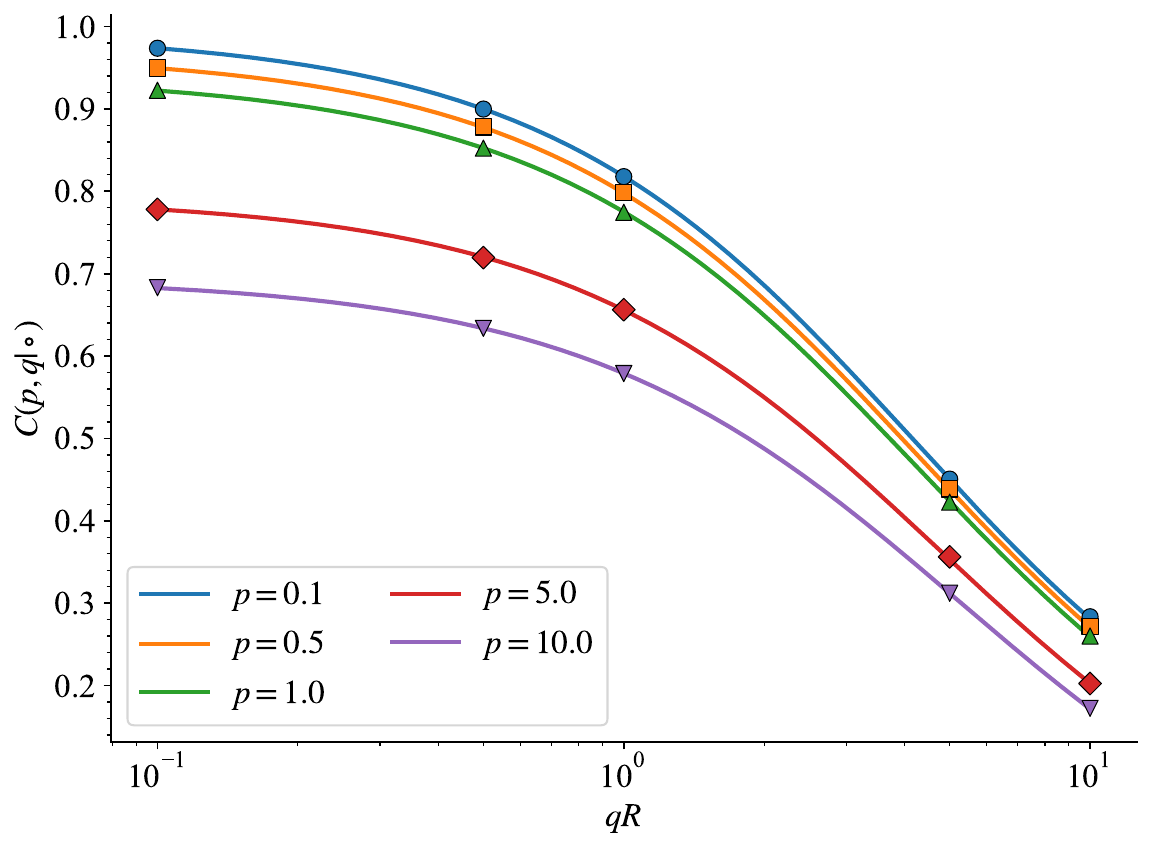}
\rlap{\hspace{-0.5\linewidth}\raisebox{-0.0\linewidth}{\textbf{(b)}}}
\caption{
The splitting probability $\phi(p,q|\circ)$ and correlation coefficient $C(p,q|\circ)$ as a function of surface-reaction rate $q$ for uniformly distributed starting points in the ball ($R=1$, $D=1$). 
Estimated by $N=10^6$ Monte Carlo trajectories, symbols correspond to different decay rates: $p=0.1$ (circles), $p=0.5$ (squares), $p=1.0$ (upward triangles), $p=5.0$ (diamonds), and $p=10.0$ (downward triangles). 
\textbf{(a)} $\phi(p,q|\circ)$. Solid curves represent the analytical result \eqref{eq:phiunif}, while colored symbols show the numerical estimates~\eqref{eq:phiest}. 
\textbf{(b)} $C(p,q|\circ)$. Solid curves represent the analytical result \eqref{eq:corr}, while colored symbols show the numerical estimates~\eqref{eq:TLtest} for Eq.~\eqref{eq:Cqdef}. 
}
\label{fig:meanapp}
\end{figure}

This appendix presents additional comparisons between the analytical predictions and Monte Carlo estimates. 
%In this section, we present several additional numerical tests to show good agreement at the resolution of the plots between analytical expressions and numerical results. 
We consider again uniformly distributed starting points in the ball  ($R=1$, $D=1$). 
Figure~\ref{fig:meanapp}(a) compares the splitting probability $\phi(p,q|\circ)$ between the analytical result~\eqref{eq:phiunif} and numerical results estimated via Eq.~\eqref{eq:phiest}, showing good agreement. For each parameter setting, $N=10^6$ Monte Carlo trajectories  were obtained.

Similarly, Fig.~\ref{fig:meanapp}(b) compares the correlation coefficient $C(p,q|\circ)$ between the analytical result~\eqref{eq:corr} and the numerical estimate for each parameter. 
The numerical correlation is calculated from the statistics of $(T_i,L_i)$ from each trajectory by Eqs.~\eqref{eq:Cqdef} and \eqref{eq:TLtest}.

\bibliography{outputOR}
%\printbibliography

\end{document}